%% file: main.tex
\documentclass[12pt]{article}

\usepackage[utf8]{inputenc}
\usepackage{amssymb}
\usepackage{amsmath}
\usepackage{subcaption}
\usepackage{graphicx}
\usepackage[margin=2.3cm]{geometry}
\usepackage{natbib}
\usepackage{xcolor}
\usepackage[official]{eurosym}
\usepackage{wrapfig}
\usepackage{booktabs}
\usepackage{graphics}
\usepackage[bookmarks=false]{hyperref}
\usepackage[graphicx]{realboxes}
\usepackage{bm}
\numberwithin{equation}{section}
\usepackage{siunitx}
\usepackage{textcomp}
\usepackage{gensymb}
\usepackage{makecell}
\usepackage[T1]{fontenc}
\usepackage{titling}
\usepackage{authblk}
\usepackage{appendix}
\usepackage{enumitem}  
\usepackage{rotating}
\usepackage{multirow}
\usepackage{amsthm}
\usepackage{algpseudocode}
\usepackage{algorithm}

\allowdisplaybreaks

\newtheorem{proposition}{Proposition}

\newtheorem{remark}{Remark}

\usepackage{orcidlink}

\title{Extremal properties of max-autoregressive moving average processes for modelling extreme river flows}
\author{Eleanor D'Arcy\orcidlink{0000-0002-7926-0758}}
\author{Jonathan A. Tawn}
\affil{\small STOR-i Centre for Doctoral Training, Department of Mathematics and Statistics, Lancaster University, LA1 4YR, UK \\ \vspace{0.2cm}\footnotesize{\textit{Correspondence to: e.darcy@lancaster.ac.uk}}}

\date{\today}

\begin{document}
\maketitle

\begin{abstract}
Max-autogressive moving average (Max-ARMA) processes are powerful tools for modelling time series data with heavy-tailed behaviour; these are a non-linear version of the popular autoregressive moving average models. River flow data typically have features of heavy tails and non-linearity, as large precipitation events cause sudden spikes in the data that then exponentially decay. Therefore, stationary Max-ARMA models are a suitable candidate for capturing the unique temporal dependence structure exhibited by river flows. This paper contributes to advancing our understanding of the extremal properties of stationary Max-ARMA processes. We detail the first approach for deriving the extremal index, the lagged asymptotic dependence coefficient, and an efficient simulation for a general Max-ARMA process. We use the extremal properties, coupled with the belief that Max-ARMA processes provide only an approximation to extreme river flow, to fit such a model which can broadly capture river flow behaviour over a high threshold. We make our inference under a reparametrisation which gives a simpler parameter space that excludes cases where any parameter is non-identifiable. We illustrate results for river flow data from the UK River Thames. 
\end{abstract}

{\it Keywords: extreme value theory, max-autoregressive moving average processes, tail dependence, extremal index, river flow extremes.}

\section{Introduction}\label{Max-ARMA:intro}
\input{Sections/intro}

\section{Model definition and parameter constraints}\label{Max-ARMA:sec:properties}
\input{Sections/definition}\label{Max-ARMA:subsec::definition}


\input{Sections/marginal}

\section{Estimating extremal measures}\label{Max-ARMA:sec:ext_measures}
\subsection{Extremal index}\label{Max-ARMA:subsec::extremalindex}
\input{Sections/extremal_index}

\subsection{Extremal dependence measure}\label{Max-ARMA:subsec::chi}
\input{Sections/chi}

\section{Simulation of Max-ARMA$(p,q)$ processes}\label{Max-ARMA:sec:simulation}
\input{Sections/sim}

\section{Inference}
\label{Max-ARMA:sec:inf}
\input{Sections/inference}

\section{Illustrative analysis of River Thames extreme flows}\label{Max-ARMA:sec:Thames}
\input{Sections/illustration}

\section{Proofs}\label{Max-ARMA:sec::proofs}
\input{Sections/proofs}

\section{Discussion}\label{Max-ARMA:sec::Discussion}
\input{Sections/discuss}


\subsection*{Acknowledgments}
This paper is based on work completed while Eleanor D'Arcy was part of the EPSRC funded STOR-i centre for doctoral training (EP/S022252/1). We would like to thank Dafni Sifnioti (EDF Energy) and Ivan Haigh (University of Southampton) for helpful discussions.
\bibliography{ref}
\bibliographystyle{apalike}

\end{document}

%% file: Sections/intro.tex
Any stationary time series that exhibits large peaks or sudden bursts of extreme observations is a candidate for modelling by a max-autoregressive moving average, Max-ARMA$(p,q)$, process for $p\in\mathbb{N}$ and $q\in\{\mathbb{N}\cup 0\}$~\citep{davis1989}. This process is a non-linear version of the well-known ARMA models~\citep{BoxJenkins1970} where a maxima replaces the summation. Max-ARMA$(p,q)$ models are suitable for data with shock noise behaviour, where the process broadly descends exponentially from each spike and then fluctuates around small values until the next major spike. 

\begin{figure}
    \centering
    \includegraphics[width=0.45\textwidth]{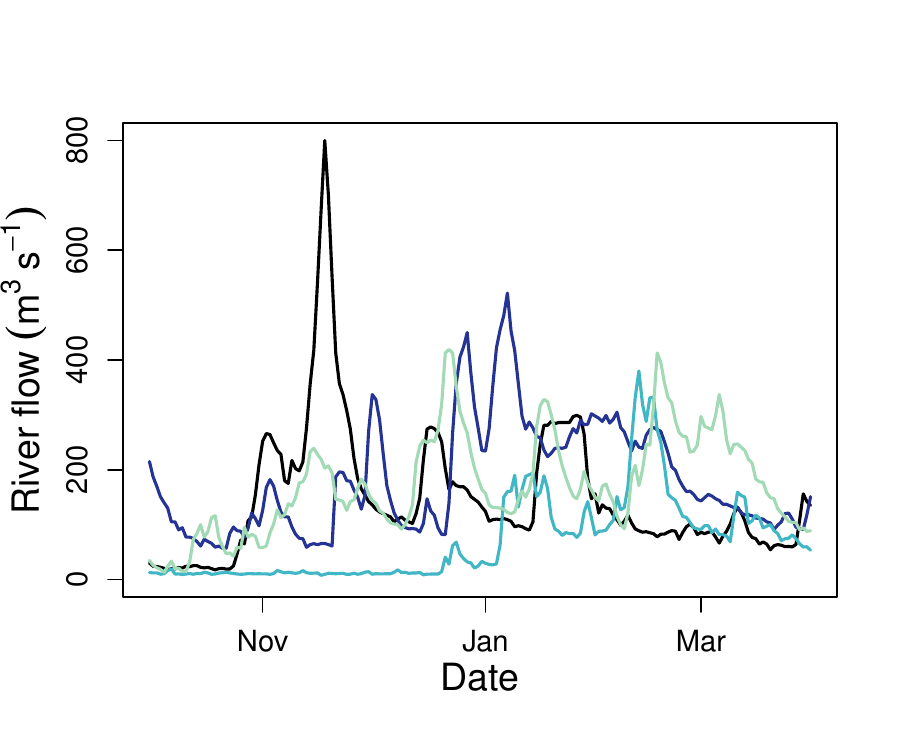}
    \includegraphics[width=0.45\textwidth]{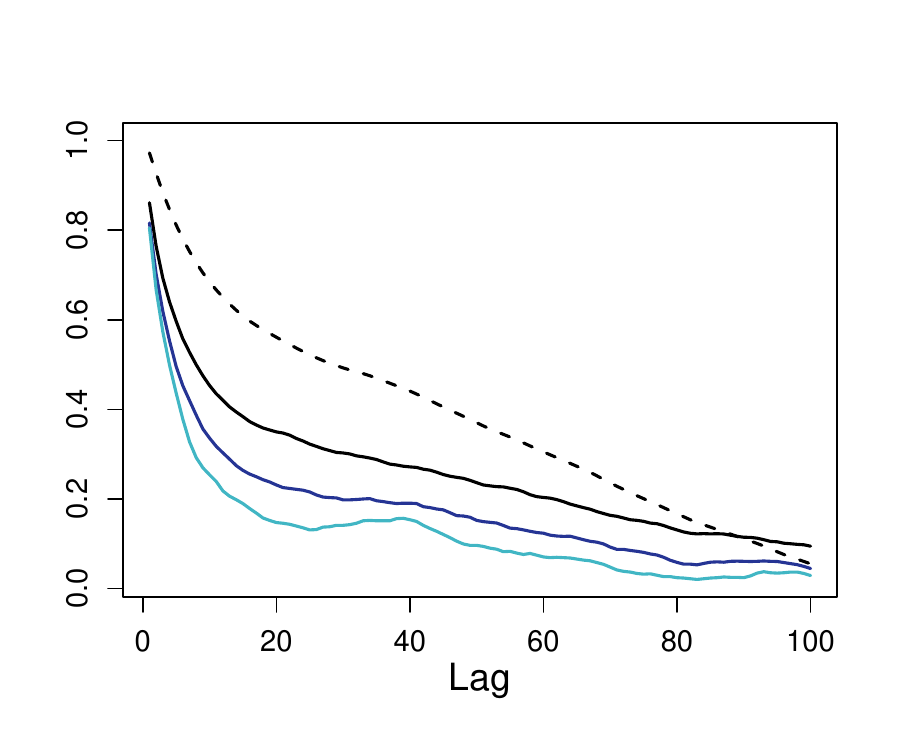}
    \caption{Left: River flow time series of the River Thames for the winter season (October - March) in 1894/95 (black), 1927/28 (dark blue), 1973/74 (light blue) and 2019/2020 (green)~\citep{EA_river_data}. Right: Pearson's correlation coefficient (dashed line) and empirical estimates of $\chi_{\kappa}(u)$ (solid lines; see Section~\ref{Max-ARMA:subsec::chi}) for the Thames data over different lags, $\kappa$ (in days). Quantiles of $u=0.9$ (black), 0.95 (dark blue) and 0.975 (light blue) are used for estimating $\chi_\kappa(u)$.}
    \label{Max-ARMA:fig:riverflow} 
\end{figure}

River flow is an example where a Max-ARMA process is a suitable candidate for modelling its behaviour, particularly in extreme states. A large precipitation event can cause river levels with sizeable catchments to remain high for days after, as the spatially distributed large volume of water takes time to propagate down the river and travel through underground systems. By adjusting the order and parameter values of the process, the frequency, width and detailed structure of the spikes can be altered. Figure~\ref{Max-ARMA:fig:riverflow} (left) shows daily maximum river flow trace plots from the River Thames (gauge at Kingston-upon-Thames) during four different winter seasons. Two of the seasons shown correspond to the worst flood events on record for the Thames: November 1894 and February 1928. The plot shows that river flow has features of heavy tails and non-linearity, with rapid rises and slower falls around the peaks. Figure~\ref{Max-ARMA:fig:riverflow} (right) shows estimates of the asymptotic dependence coefficients $\chi_{\kappa}$ (see Section~\ref{Max-ARMA:subsec::chi}) at different lags $\kappa\in[1,100]$, which provides a similar type of information as for an auto-correlation function, but here with a metric to measure extremal dependence across time. The estimated $\chi_{\kappa}$ values
demonstrate strong temporal dependence until approximately lag 14, corresponding to 2 weeks, after which the dependence steadily decays. Estimates of $\chi_{\kappa}$ vary with the threshold used to define an extreme event, with illustrations of estimates using three possible thresholds (see Section~\ref{Max-ARMA:sec:simulation} for further discussion of these estimates). Figure~\ref{Max-ARMA:fig:riverflow} (right) also shows Pearson's correlation coefficient over time lags $\kappa$ to illustrate the differing dependence structure in the body and tail of the data; dependence is stronger in the body until approximately lag 90. Similar to the estimates of $\chi_\kappa$, there is a change in the rate that the dependence measure decays beyond lag 14, after which the correlation coefficient decays at a steadier rate than for $\kappa\leq14$, although the decay is much quicker than for $\chi_\kappa$ across all lags.

Joint modelling of sea levels and river flow is fundamental for forecasting future closures of flood barriers, such as the Thames Barrier, in both the short- and long-term. Short-term predictions are fundamental for anticipating closure times and durations to ensure preparedness~\citep{dale2014, ThamesBarrier}. We are interested in longer-term forecasts to explore the impacts of climate change on the number of barrier closures. Increases in barrier closures, in turn, require an increase in the number of maintenance and safety checks. This affects the reliability of the barrier because during these checks, the barrier cannot be used and, due to time constraints, the number of checks per year is restricted. Under climate change conditions, the number of times the barrier can close each year is likely to be exceeded soon~\citep{ThamesBarrier}. Further measures must be taken to ensure the barrier can remain in operation, but if this is not possible, to plan for barrier updates or replacement. Therefore, by forecasting changes in future closure rates in the long term, one can provide critical information to barrier operators about barrier management and provide insight into its life expectancy~\citep{tracekleeberg2023}.

The barrier closes when sea levels and/or river flows exceed pre-determined, confidential, thresholds. Therefore, it is fundamental that we focus on modelling the extremes of both variables as these values are more likely to contribute to barrier closures. A novel method for estimating extreme sea levels was recently developed by~\cite{DArcy2023} for tide gauge data at Sheerness, located at the mouth of the River Thames. Using a Max-ARMA($p,q$) process, our aim is to develop a similar method for estimating extreme river flow. Simulations for river flow and sea levels can then be combined to determine closure rates, as these variables are found to be approximately independent~\citep{SvenssonJones2004}. Of course, the Max-ARMA($p,q$) model for river flow is not restricted to flood barrier and estuary locations, but can be applied to all river flow data in no-drought areas.

Processes with Max-ARMA behaviour exhibit unique temporal dependence structures that we aim to explore at extreme levels.~\cite{davis1989} derive conditions on the parameters of a Max-ARMA($p,q$) process for it to be stationary and under which each of the parameters of the model is identifiable. We derive further conditions for all parameters to be identifiable and present a simplified parameter space that is useful for inference. We are interested in deriving the well-known extremal index $\theta$~\citep{Leadbetter1983} and the coefficient of asymptotic dependence $\chi_\kappa$ at lags $\kappa\in\mathbb{Z}$; the latter is a special case of the extremogram defined by~\cite{DavisMikosch2009}. The Max-ARMA(1,0) process has been studied previously:~\cite{robinsontawn2000} and~\cite{Ferreira2011} derived the form of the $\theta$ and $\chi_\kappa$, respectively. However, to the best of our knowledge, these measures have not yet been derived for a general Max-ARMA($p,q$) process. We illustrate how these measures can be used for moments-based inference using the River Thames data as an example.

Max-ARMA($p,q$) processes are defined on specific margins, for example, we use unit Fr\'echet margins. This can be thought of as in copula theory where the margins can be changed using the probability integral transform~\citep{nelsen2006}. Therefore we need an additional transformation from the observed series $\{Y_t\}$ to the Max-ARMA($p,q$) series $\{X_t\}$, via $X_t=T(Y_t)$, for all $t\in\mathbb{Z}$, where $T$ is defined in Section~\ref{Max-ARMA:subsec:marg_inf}. This transformation allows us to compare the observed series in their extreme states, such as those in Figure~\ref{Max-ARMA:fig:riverflow}, with simulated Max-ARMA($p,q$) realisations.

In Section~\ref{Max-ARMA:sec:properties} we formally define the Max-ARMA($p,q$) process and derive conditions on its parameters for this to be an identifiable process. In Section~\ref{Max-ARMA:sec:ext_measures} we detail the form of the extremal index and the coefficient of asymptotic dependence. In Section~\ref{Max-ARMA:sec:simulation} we explain how to simulate from a Max-ARMA process and provide examples of such simulations.~\cite{davis1989} find super-efficient estimators of the parameters for known $(p,q)$; in Section~\ref{Max-ARMA:sec:inf} we take a different inference approach, as we believe that the Max-ARMA$(p,q)$ process is only an approximation to the true generating process (of river flows) and when they are in an extreme state. We propose a generalised moments-based inference procedure for the joint behaviour of exceedances of a high threshold. In Section~\ref{Max-ARMA:sec:Thames} we illustrate this inference method on data from the River Thames. All proofs are given in Section~\ref{Max-ARMA:sec::proofs}. We conclude with a discussion in Section~\ref{Max-ARMA:sec::Discussion}.

%% file: Sections/definition.tex
\cite{davis1989} define a discrete-time stochastic process $\{X_{t}\}$ for $-\infty<t<\infty$ as following a Max-ARMA$(p,q)$ model if, \begin{equation}
    X_{t}=\max\{\alpha_1 X_{t-1}, \ldots, \alpha_{p}X_{t-p}, \beta_0 Z_{t}, \beta_1 Z_{t-1}, \ldots, \beta_q Z_{t-q}\},\label{Max-ARMA:eq::Max-ARMA_definition}
\end{equation} 
where 
$\bm\alpha=(\alpha_1, \ldots, \alpha_p)$ and 
$\bm\beta=(\beta_0, \ldots, \beta_q)$ are parameters of the model, for $p\in\mathbb{N}$ and $q\in\{0,\mathbb{N}\}$, and $\{Z_t\}$ is an independent and identically distributed (IID) innovation process on $\text{Fr\'echet}(\gamma)$ margins, so that $F_{Z}(z)=\exp(-\gamma/z)$ for $z>0$ and $\gamma>0$. For the model to be well-defined the parameters must satisfy the constraints
$\alpha_i \geq 0$ for $i=1,\ldots,p-1$ and $\alpha_p>0$, $\beta_0=1$, $\beta_j \geq 0$ for $j=1,\ldots,q-1$ and $\beta_q>0$. Despite the restriction on $\beta_0$, we use the more general $\beta_0$ notation throughout
for mathematical convenience for expressing results. By construction~\eqref{Max-ARMA:eq::Max-ARMA_definition},
$X_t$ and $Z_s$ are independent for all $s>t$.

%% file: Sections/marginal.tex
We make further restrictions on the parameters $\alpha_i$ for $i=1,\ldots,p$ to ensure the process is stationary and derive the form of $\gamma$, the scale parameter for the innovation process, such that $\{X_t\}$ has a specific marginal form, in the remark below.

\begin{remark}\label{Max-ARMA:rmk::gamma}
A Max-ARMA($p,q$) process is a stationary process with Fr\'echet($\sigma$) marginal distribution if and only if $0\leq \alpha_i < 1$ for $i=1,\ldots,p-1$ and $0 < \alpha_p < 1$, and the scale parameter $\gamma>0$ of the Fr\'echet distribution for the innovation process $\{Z_t\}$ is given by \begin{equation}
     \gamma := \sigma \Bigg(\sum\limits_{\tau=0}^{\infty}\max_{\mathcal{S}_\tau}\bigg\{ \beta_j\prod_{\substack{i=1,\ldots,p:\\\alpha_i>0}}\alpha_i^{a_i}\bigg\}\Bigg)^{-1} < \infty,\label{Max-ARMA:eq::gamma}
\end{equation} 
for $0<\sigma<\infty$, where \begin{equation}
    \mathcal{S}_\tau=\bigg\{ (i,j,a_i)\in\{1,\ldots,p\}\times\{0,\ldots,q\}\times\{0,1,\ldots,\tau\}:\sum_{i=1}^p ia_i + j= \tau\bigg\}. \label{Max-ARMA:eq::set}
\end{equation}
\end{remark}

\noindent Remark~\ref{Max-ARMA:rmk::gamma} is a reformulation of the result in~\cite{davis1989} into a simpler expression, and we explicitly state the marginal distribution for $\{X_t\}$. The formulation of~\cite{davis1989} means that the Max-ARMA parameters determine both marginal and dependence parameters, whereas our standardisation of the margins, through the choice of $\gamma$, makes the dependence parameters independent of the marginal distribution, which is particularly useful for copula style inferences, which is required in Section~\ref{Max-ARMA:subsec:inference}. Stationarity requires no further restrictions on $\bm\beta$. From this point onwards in our theoretical developments, we take $\sigma=1$.

For a Max-ARMA(1,0) process, defined by $X_t=\max\{\alpha_1 X_{t-1}, Z_{t}\}$, to have unit Fr\'echet margins with stationary constraints given in Remark~\ref{Max-ARMA:rmk::gamma}, i.e., $0< \alpha_1 < 1$, we obtain that $0<\gamma< 1$, with
\begin{equation}
\gamma=\Bigg(\sum_{\tau=0}^{\infty}\alpha_1^\tau\Bigg)^{-1}=1-\alpha_1.
\label{Max-ARMA:eq:gamma_Max-ARMA10}
\end{equation} 

Next, we impose further restrictions on $(\bm\alpha, \bm\beta)$, to ensure that the parameters are identifiable, i.e., in the sense that they have some effect on the feasible sample paths of the $\{X_t\}$ process. Remark~\ref{Max-ARMA:rmk:identifiable} presents
a result by~\cite{davis1989}, with a minor adaptation to allow for elements of $\bm\alpha$ to be zero. 
\begin{remark}
For a stationary Max-ARMA($p,q$) process with $p\geq 2$, $\alpha_k$ is identifiable only if 
\begin{equation*}
    \max_{\mathcal{R}_k}\bigg\{\prod_{\substack{i=1,\ldots,(k-1):\\\alpha_i>0}}\alpha_i^{a_i} \bigg\} < \alpha_k<1,
\end{equation*} for $k=2,\ldots,p$, where
\[
\mathcal{R}_k = \{\;(i,a_i)\in\{1,\ldots,(k-1)\}\times\{0,1,\ldots,k\}: \sum_{i=1}^{k-1} ia_i = k\}.
\]
 Since $\beta_0=1$, it is always identifiable. For $j\in 1,\ldots,\min\{p,q\}$, $\beta_j$ is identifiable if $\beta_j > \alpha_j$; and when 
 $q>p$ if  $\beta_j>0$  it is identifiable if $j\in \{p+1,\ldots,q\}$. 
\label{Max-ARMA:rmk:identifiable} 
\end{remark}

Note that $\mathcal{R}_k$ is defined similarly to $\mathcal{S}_\tau$ in expression~\eqref{Max-ARMA:eq::set} for a Max-ARMA$(k-1,0)$ process, so that there are no $\bm\beta$ parameters. In the following proposition, we extend Remark~\ref{Max-ARMA:rmk:identifiable}, which gives a condition for the marginal identifiability of $\alpha_i$, to derive a condition under which all $\alpha_i$ for $i=1,\ldots,p$ are identifiable. We prove this result in Section~\ref{Max-ARMA:sec::proofs}.

\begin{proposition}
    For a stationary Max-ARMA($p,q$) process with $p\geq 2$, if $\alpha_i$ for all $i=1,\ldots,(k-1)<p$ are identifiable, then $\alpha_k$ is identifiable if 
    \begin{eqnarray}
        \max_{i=1,\ldots ,\lfloor k/2\rfloor} \{\alpha_i\alpha_{k-i}\} & < & \alpha_k,
        \label{Max-ARMA:eqn:alphalowerbound}
    \end{eqnarray} 
    where $\lfloor\cdot\rfloor$ denotes the floor function. Therefore, all $\alpha_i$ for $i=1,\ldots,p$ are identifiable if the above holds for all $i$.
\label{Max-ARMA:prop:identifiability}
\end{proposition}

\begin{remark}
All parameters $\alpha_2,\ldots,\alpha_p$ are not identifiable if $\alpha_i\leq\alpha_1^{i}$ for $i=2,\ldots,p$. 
\end{remark}

\begin{remark}
If $\alpha_1=0$ and $p\ge 3$, then by Proposition~\ref{Max-ARMA:prop:identifiability}, if $\min\{\alpha_2,\alpha_3\}>0$, both $\alpha_2$ and $\alpha_3$  are identifiable. However, for all remaining $\bm\alpha$ parameters to be identifiable, Proposition~\ref{Max-ARMA:prop:identifiability} must hold, with this simplifying, 
for all $i=4, \ldots ,p$, to 
\begin{eqnarray*}
        \max_{j=2,\ldots ,\lfloor i/2\rfloor} \{\alpha_j\alpha_{i-j}\} & < & \alpha_i.
\end{eqnarray*}
\label{Max-ARMA:rmk:notidentifiable} 
\end{remark}
In Section~\ref{Max-ARMA:sec:Thames} we discuss the implications of imposing these stationarity and identifiability constraints on parameter inference for the Max-ARMA process.

%% file: Sections/extremal_index.tex
Consider the maximum $M_n$ of a sequence of IID unit Fr\'echet random variables $X_1,\ldots,X_n$, i.e., $M_n=\max\{X_1,\ldots,X_n\}$. If there exist sequences of constants $\{c_n>0\}$ and $\{d_n\}$, so that the rescaled block maximum $(M_n-d_n)/c_n$ has a nondegenerate limiting distribution $G$ as $n\rightarrow\infty$, then $G$ is a member of the generalised extreme value distribution (GEV) family~\citep{Leadbetter1983}. In this case, for unit Fr\'echet random variables, using $c_n=n$ and $d_n=0$, $G$ is also unit Fr\'echet so that $G(x)=\exp(-1/x)$ for $x>0$; this is a member of the GEV family.

Now consider $X_1,\ldots,X_n$ as a stationary sequence, still with unit Fr\'echet margins, but satisfying the long-range dependence condition, $D(u_n)$ of~\cite{Leadbetter1983} with $u_n=c_n x+d_n$ for any $x$ in the domain of $G$, which ensures events long apart in time are near independent but does not impose any bounds on the local dependence conditions. The limiting distribution of $(M_n-d_n)/c_n$ of this stationary process when it exists, with $c_n$ and $d_n$ as for the IID case, is $G^{\theta}(x)$ with $G(x)=\exp(-1/x)$ for $x>0$ and $\theta\in(0,1]$ the extremal index. For an independent series $\theta=1$, but the converse is not true, so $\theta$ captures the effect of the temporal dependence on the distribution of the maximum.  

For dependent sequences, clusters of extreme events form above high thresholds. Extreme observations in different clusters are assumed to be independent, whilst events within the same cluster exhibit dependence. Let $N_{p_n}(u_n)$ denote the number of observations of $X_1,\ldots,X_{p_n}$ to exceed the threshold $u_n$, where $p_n=\mathcal{O}(n)$. Then a cluster occurs when $N_{p_n}(u_n)>0$, so the cluster size distribution $\pi$ is defined by \begin{equation*}
    \pi(m)=\lim_{n\rightarrow\infty}\Pr(N_{p_n}(u_n)=m|N_{p_n}(u_n)>0),
\end{equation*} for $m\in\mathbb{N}$~\citep{hsing1988exceedance}. The extremal index, defined by~\cite{Leadbetter1983}, is characterised as the reciprocal of the limiting mean cluster size, where \textit{limiting} refers to exceedances of an increasing threshold tending to the upper endpoint. \cite{obrien1987} gives an alternative definition of the extremal index, based on the number of down-crossings in clusters of threshold exceedances, i.e., \begin{equation}
    \theta = \lim_{n\rightarrow\infty}\Pr(X_{2}\leq u_n, \ldots, X_{p_n}\leq u_n | X_{1} > u_n),
    \label{Max-ARMA:eqn:OBrien}
\end{equation} where $\{p_n\}$ is an increasing sequence with $p_n=\mathcal{O}(n)$. They give an alternative long-range dependence condition to that of~\cite{Leadbetter1983}, known as the asymptotic independence of maxima condition. The well-known runs method of estimation for the extremal index~\citep{SmithWeissman1994} is strongly related~\citeauthor{obrien1987}'s definition, where clusters are defined as those separated by $p_n$ non-extreme values, where $p_n$ is termed the run length. 

For a Max-ARMA(1,0) process, \cite{robinsontawn2000} show that $\theta=1-\alpha_1$. They also show that for this particular process, the cluster sizes are geometrically distributed, with distribution function $\pi(m) = \alpha_1^{m-1}(1-\alpha_1)$ for $m\in\mathbb{N}$. This is a suitable model as the number of exceedances determines a cluster, whilst a non-exceedance would terminate a cluster. To the best of our knowledge, the extremal index of a Max-ARMA$(p,q)$ process has not been derived before. We prove the following result in Section~\ref{Max-ARMA:sec::proofs}.

\begin{proposition}\label{Max-ARMA:prop::extremalindex}
    The extremal index of a stationary Max-ARMA$(p,q)$ process is \begin{equation}
        \theta = \gamma\max\{\beta_0,\beta_1,\ldots,\beta_q\},\label{Max-ARMA:eq:theta}
    \end{equation} where $\gamma>0$ is the scale parameter of the innovation $Z_t$ distribution, defined by expression~\eqref{Max-ARMA:eq::gamma}. 
\end{proposition}
As $\beta_0=1$, then $\max\{\beta_0, \ldots ,\beta_q\}\ge 1>\max\{\alpha_1, \ldots ,\alpha_p\}$ due to stationarity. Hence if 
$\beta_J=\max\{\beta_0, \ldots ,\beta_q\}$, then 
$\beta_J>\alpha_J$ and hence it is identifiable by Remark~\ref{Max-ARMA:rmk:identifiable}, so the result in Proposition~\ref{Max-ARMA:prop::extremalindex} is well defined. According to the above proposition, a Max-ARMA(1,0) process with $\beta_j=0$ for all $j$ and $\gamma=1-\alpha_1$, as identified by expression~\eqref{Max-ARMA:eq:gamma_Max-ARMA10}, has extremal index $\theta=1-\alpha_1$. This agrees with the results of~\cite{robinsontawn2000}. 

%% file: Sections/chi.tex
An alternative measure of extreme temporal dependence to the extremal index is the coefficient of extremal dependence $\chi_\kappa$ for specific lags $\kappa$~\citep{Coles1999, heffernan2007}. When used for a stationary time series context, this is a similar type of pairwise dependence measure to the auto-correlation function used for classic ARMA models. This is defined as \begin{equation}
    \chi_\kappa=\lim_{u\rightarrow x^F}\Pr(X_{t+\kappa}>u|X_{t}>u),
    \label{Max-ARMA:eqn:ChiTau}
\end{equation} 
for $k\in\mathbb{Z}$ and a threshold $u$, where $x^F$ is the upper endpoint of the marginal distribution function $F_X$, so for a Max-ARMA process with Fr\'echet margins $x^F=+\infty$. If $\chi_\kappa\in(0,1]$, we say that $X_t$ and $X_{t+\kappa}$ are asymptotically dependent; this means there is a non-zero probability of $X_{t+\kappa}$ being large when $X_{t}$ is large at all extreme levels. Whereas $\chi_\kappa=1$ and $\chi_\kappa=0$ correspond to perfect dependence in the extremes and asymptotic independence, respectively. \cite{Ferreira2011} derives $\chi_\kappa$ for the Max-ARMA(1,0) process with Pareto($\xi_P$) margins, as $\chi_\kappa = \alpha_1^{\mid \kappa/\xi_P\mid}$ where $\xi_P>0$ is the tail index of the Pareto distribution. Here, we derive $\chi_\kappa$ for a general Max-ARMA($p,q$) process which has not been done before, to the best of our knowledge.

A similar measure of serial extremal dependence is the extremogram, defined by~\cite{DavisMikosch2009}. For a regularly varying stationary series $\{X_t\}$, the extremogram is defined for two sets $A$ and $B$, that are bounded away from zero, by the limit\begin{equation*}
    \rho_{AB}(\kappa)=\lim_{n\rightarrow\infty}\mathbb{P}(c_n^{-1}X_{t}\in{A}\;|\;c_n^{-1}X_{t+\kappa}\in{B}),
\end{equation*} for $c_n\rightarrow\infty$ such that $\mathbb{P}(|X_t|>c_n)\sim{n^{-1}}$. By defining $A$ and $B$ as bounded away from zero, the events $\{c_n^{-1}X_{t}\in{A}\}$ and $\{c_n^{-1}X_{t+\kappa}\in{B}\}$ become extreme in the limit. The extremogram becomes the extremal dependence measure above when $A=B=(1,\infty)$ and $c_n=n$, with the expression of $c_n$ given by $\{X_t\}$ having Fr\'echet margins. 

For $\tau\in\mathbb{N}$ we define \begin{equation}
  \gamma_{\tau}:=\max_{\mathcal{S}_\tau} \Bigg\{\beta_j\prod_{\substack{i=1,\ldots,p:\\\alpha_i>0}}\alpha_i^{a_i}\Bigg\},
    \label{Max-ARMA:eqn:Edefn}
    \end{equation}
where $\mathcal{S}_{\tau}$ is defined by expression~\eqref{Max-ARMA:eq::set}. 
Then $0 \le \gamma_\tau \le \max\{\beta_0, \beta_1,\ldots,\beta_q\}$. If $\max\{\beta_1,\ldots,\beta_q\}\\>1$, then $\gamma_\tau>1$ for some $\tau$ but $\gamma_\tau\rightarrow0$ always as $\tau\rightarrow\infty$ for a stationary Max-ARMA process, since $\max\{\alpha_1, \ldots ,\alpha_p\}<1$ and for each $\tau$ there exist is at least one $i=1,\ldots ,p$ such that $a_i\rightarrow\infty$ as $\tau\rightarrow\infty$. This gives us the final notation required for the following proposition about $\chi_k$ for a Max-ARMA$(p,q)$ process, which we prove in Section~\ref{Max-ARMA:sec::proofs}.

\begin{proposition}\label{Max-ARMA:prop::chi}
    The extremal dependence measure $\chi_\kappa$ for observations separated by $\kappa\in\mathbb{N}$ time lags, for a stationary Max-ARMA$(p,q)$ process $\{X_t\}$, is \begin{equation}
        \chi_\kappa =  \gamma\sum\limits_{\delta=0}^{\infty}\min\{\gamma_\delta,\gamma_{\delta+\kappa}\},\label{Max-ARMA:eq:chi}
    \end{equation} 
    where $\gamma_\delta$ is defined by expression~\eqref{Max-ARMA:eqn:Edefn}, and $\chi_{-\kappa}=\chi_{\kappa}$. 
\end{proposition}

\begin{remark}
It follows from the formulation of $\gamma_{\tau}$ in expression~\eqref{Max-ARMA:eqn:Edefn} that $\gamma_{(q+1)\tau+j}$ decays at a geometric rate for each $j=1,\ldots ,q$ each converging  to zero as $\tau\rightarrow \infty$. So $\lim_{\kappa\rightarrow \infty} \chi_{\kappa}=0$.
\end{remark}

\begin{remark}
When $1>\alpha_1>\ldots >\alpha_p>0$ and $\beta_0=1>\beta_1>\ldots>\beta_q>0$, $\gamma_\tau$ is a strictly decreasing function for all $\tau\in\mathbb{N}$. Then the form of $\chi_\kappa$ for $\kappa\ge 0$ simplifies as follows, 
\begin{equation*}
\chi_{\kappa} = \gamma\sum\limits_{\tau=0}^{\infty}\gamma_{\tau+\kappa}=
\gamma\sum\limits_{\tau=\kappa}^{\infty}\gamma_{\tau} = \gamma\left(
\sum\limits_{\tau=0}^{\infty}\gamma_{\tau}-\sum\limits_{\tau=0}^{\kappa -1}\gamma_{\tau} \right)= \gamma\left(\frac{1}{\gamma}- \sum\limits_{\tau=0}^{\kappa -1}\gamma_{\tau}\right)= 1 -\gamma 
\sum\limits_{\tau=0}^{\kappa -1}\gamma_{\tau}, 
\end{equation*} where the infinite sum is equal to $1/\gamma$ by Remark~\ref{Max-ARMA:rmk::gamma}.\label{Max-ARMA:rmk:simplify_chi}
\end{remark}

For the stationary Max-ARMA(1,0) process, $\gamma=1-\alpha_1$ from expression~\eqref{Max-ARMA:eq::gamma}, and as $\beta_0=1$, $\gamma_{\tau}=\alpha_1^{\tau}$ from expression~\eqref{Max-ARMA:eqn:Edefn} so, for $\kappa\geq 0$, we obtain that  
\begin{equation*}
    \chi_\kappa= 1-(1-\alpha_1)\sum\limits_{\delta=0}^{\kappa-1}\alpha_1^\delta=\alpha_1^{\kappa},
\end{equation*}
which agrees with the result of \cite{robinsontawn2000}.

%% file: Sections/sim.tex
\cite{davis1989} simulate from a stationary Max-ARMA(1,0) process, which is straight-forward as $X_{1}$ can be generated from the marginal distribution. and then expression~\eqref{Max-ARMA:eq::Max-ARMA_definition} can be used recursively for $t>2$. However, they are not specific about how to generate $X_1$, but calculating $\gamma$ in this case is straightforward using expression~\eqref{Max-ARMA:eq:gamma_Max-ARMA10}. Extending this method to a higher order, particularly with $q\ge 1$ is more complex; we illustrate how to do so in this section.

We aim to simulate a series of length $n\in\mathbb{N}$, denoted by $\{\tilde{X}_t\}_{t=1}^{n}$, from a stationary Max-ARMA($p,q$) process. Since the Max-ARMA process is driven by an IID innovation process $Z_{t}\sim\text{Fr\'echet}(\gamma)$, we first simulate $m+n$ values from this distribution, denoted by $\{\tilde{Z}_t\}_{t=-m}^n$, with scale parameter $\gamma>0$ determined by the parameters of the Max-ARMA process, as stated in Remark~\ref{Max-ARMA:rmk::gamma}, to ensure stationarity. Here, we have $m$ values linked to the requirement for a burn-in period; see below for the reasons why. 

The first step is to derive the value of $\gamma$, which is given by an infinite sum~\eqref{Max-ARMA:eq::gamma} of terms, where the $\tau$th  term itself is the maximum over a complex set $\mathcal{S}_{\tau}$. Given parameter values $\bm\alpha$ and $\bm\beta$, we approximate the infinite sum by a partial sum up to the $N$th term, for large $N$. We find that $N=100$ gives a suitable approximation for a wide range of $\bm\alpha$ and $\bm\beta$, but larger values may be necessary when $\max\{\alpha_1,\ldots,\alpha_p\}$ is very close to 1.

For each step $\tau=0,\ldots,N$ there is a need to find a solution set that satisfies the conditions of $\mathcal{S}_{\tau}$. We achieve this by trying all possible combinations and keeping only the feasible set. Specifically, we check for all possible combinations of $a_i\in\{0,\ldots,\tau:\alpha_i>0\}$ for $i=1,\ldots,p$ and $j\in\{0, 1,\ldots,q\}$ that satisfy the properties of $\mathcal{S}_\tau$ outlined in~\eqref{Max-ARMA:eq::set}, for each iteration of $\tau$. We then find which element of this feasible solution set maximises $\beta_j\prod_{i=1,\ldots,p\;:\;\alpha_i>0}\alpha_i^{a_i}$.

First, consider how to simulate forward given the process is already in a stationary state. At time $t\in \mathbb{N}$, with $p\leq t \leq n$, we can easily simulate $\tilde X_t$ using expression~\eqref{Max-ARMA:eq::Max-ARMA_definition} since the {\it past values} $(\tilde{X}_{t-1}, \ldots ,\tilde{X}_{t-p})$ and $(\tilde{Z}_{t-1}, \ldots ,\tilde{Z}_{t-q})$ are available. However, for the earliest terms in the simulated $\tilde{X}_t$ sequence where $t<p$, the {\it past values} are unknown so the joint distribution for $\tilde X_t$ is complex. Instead, we simulate these early terms $\tilde{X}_1,\ldots,\tilde{X}_p$ from a unit Fr\'echet distribution. These observations have the correct marginal distribution but are independent of one another. Therefore a burn-in period of length $m\geq q-\min\{p,q\}$ is required to ensure that the simulated series has the correct stationary dependence structure.

We illustrate the extremal properties of a stationary Max-ARMA$(p,q)$ process through four examples, labelled series 1-4, where a burn-in period of length $m=1000$ is discarded. Series 1 and 2 are Max-ARMA(3,0) processes whilst series 3 and 4 are Max-ARMA(3,3) processes. The values of $(\bm\alpha, \bm\beta)$ are given in Table~\ref{Max-ARMA:tab:sim_results} and are chosen to ensure identifiability, except for series 2 where we explore the effect of $\alpha_2=0$ with $\alpha_1,\alpha_3$ identifiable. This table also presents the values of $\gamma$, $\theta$ and $\chi_\kappa$ ($\kappa=1,2,3$) for each series, where these are evaluated using the limiting theoretical expressions given by Remark~\ref{Max-ARMA:rmk::gamma} and Propositions~\ref{Max-ARMA:prop::extremalindex} and~\ref{Max-ARMA:prop::chi}, respectively. Series 2 exhibits less extremal dependence than the remaining series due to its smaller $\bm\alpha$ parameters. Series 1, 3 and 4 have similar values for $\theta$ and $\chi_\kappa$; this shows that adding $\bm\beta$ parameters to series 1 hasn't changed the extremal dependence structure and neither has increasing the magnitude of the $\bm\beta$ parameters, as in series 4. The values of $\theta$ indicate a cluster size of 9 for processes 1, 3 and 4. The values of $\chi_\kappa$ decay with increasing lag $\kappa$ in all cases.

\begin{table}[t]
    \centering
    \resizebox{\linewidth}{!}{
    \begin{tabular}{ccccccccc}
    \hline
    \multicolumn{5}{c}{Series} & \multicolumn{4}{c}{Extremal properties} \\
    & $(p,q)$ & $\bm\alpha$ & $\bm\beta$ & $\gamma$ & $\theta$ & $\chi_1$ & $\chi_2$ & $\chi_3$ \\  \hline
    1 & $(3,0)$ & $(0.85,0.77,0.7)$ & $(0,0,0)$ & 0.11 & 0.11 (0.11) & 0.88 (0.88) & 0.79 (0.80) & 0.70 (0.71) \\
    2 & $(3,0)$ & $(0.3,0,0.1)$ & $(0,0,0)$ & 0.65 & 0.65 (0.58) & 0.35 (0.36) & 0.16 (0.19) & 0.1 (0.14) \\
    3 & $(3,3)$ & $(0.85,0.77,0.7)$ & $(2,1,0.9)$ & 0.05 & 0.11 (0.10) & 0.89 (0.88) & 0.8 (0.79) & 0.72 (0.72) \\
    4 & $(3,3)$ & $(0.85,0.77,0.7)$ & $(50,10,5)$ & 0.002 & 0.11 (0.11) & 0.89 (0.87) & 0.79 (0.78) & 0.70 (0.70) \\
    \hline
    \end{tabular}
    }
    \caption{Values of $\gamma$, $\theta$ and $\chi_\kappa$ for $\kappa=1,2,3$ of different Max-ARMA processes derived from Remark~\ref{Max-ARMA:rmk::gamma} and Propositions~\ref{Max-ARMA:prop::extremalindex} and~\ref{Max-ARMA:prop::chi}, respectively. Empirical estimates of $\theta(u)$ and $\chi_\kappa(u)$, derived using a simulation of length $10^6$ and the $0.95$ marginal quantile as the threshold $u$, are given in parentheses. All values are given to 2 decimal places, where appropriate.}
    \label{Max-ARMA:tab:sim_results}
\end{table}

\begin{figure}
    \centering
    \includegraphics[width=0.49\textwidth]{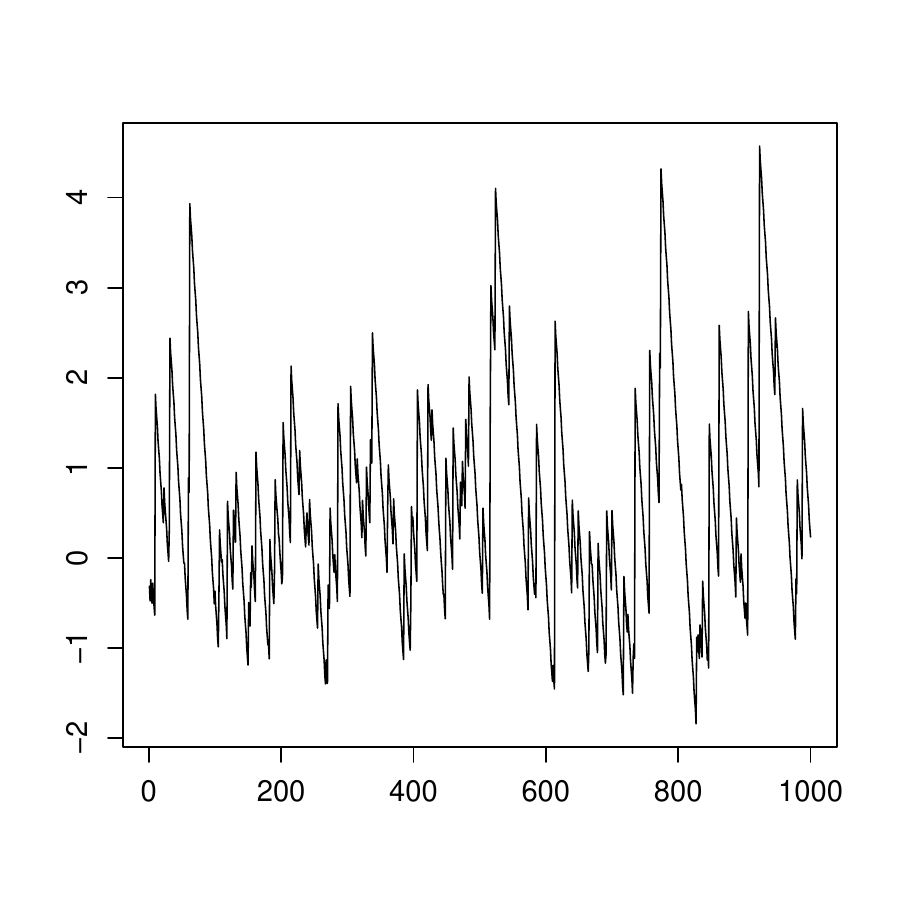}
    \includegraphics[width=0.49\textwidth]{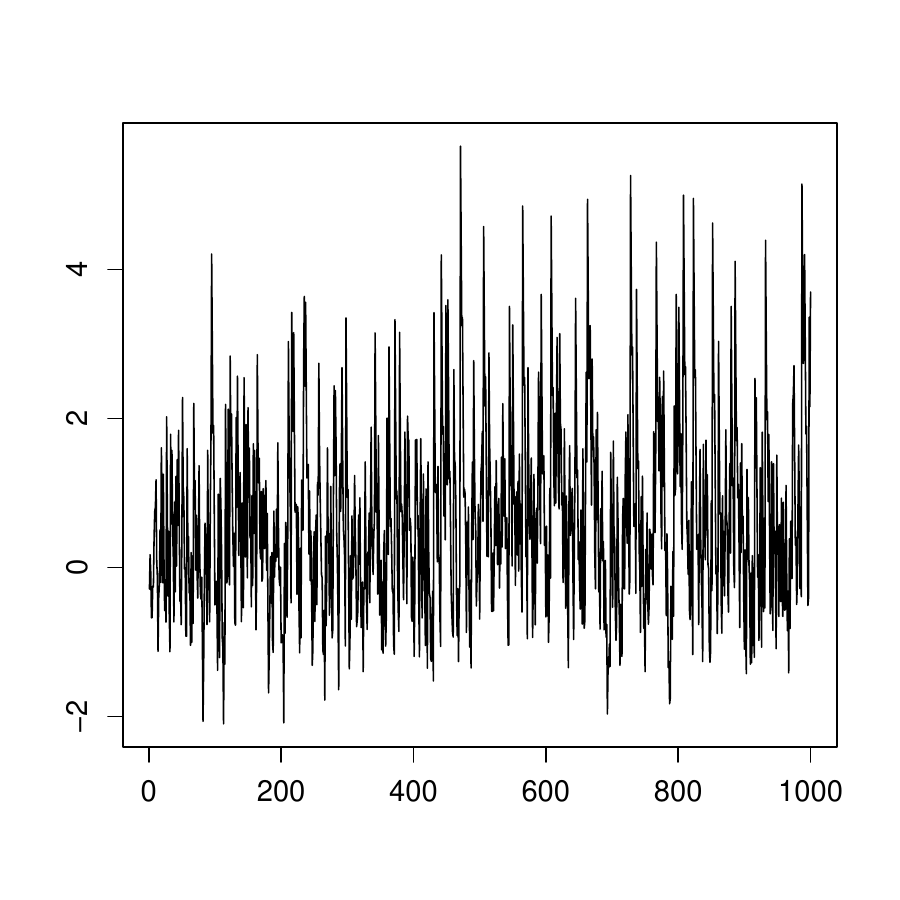}\\
    \includegraphics[width=0.49\textwidth]{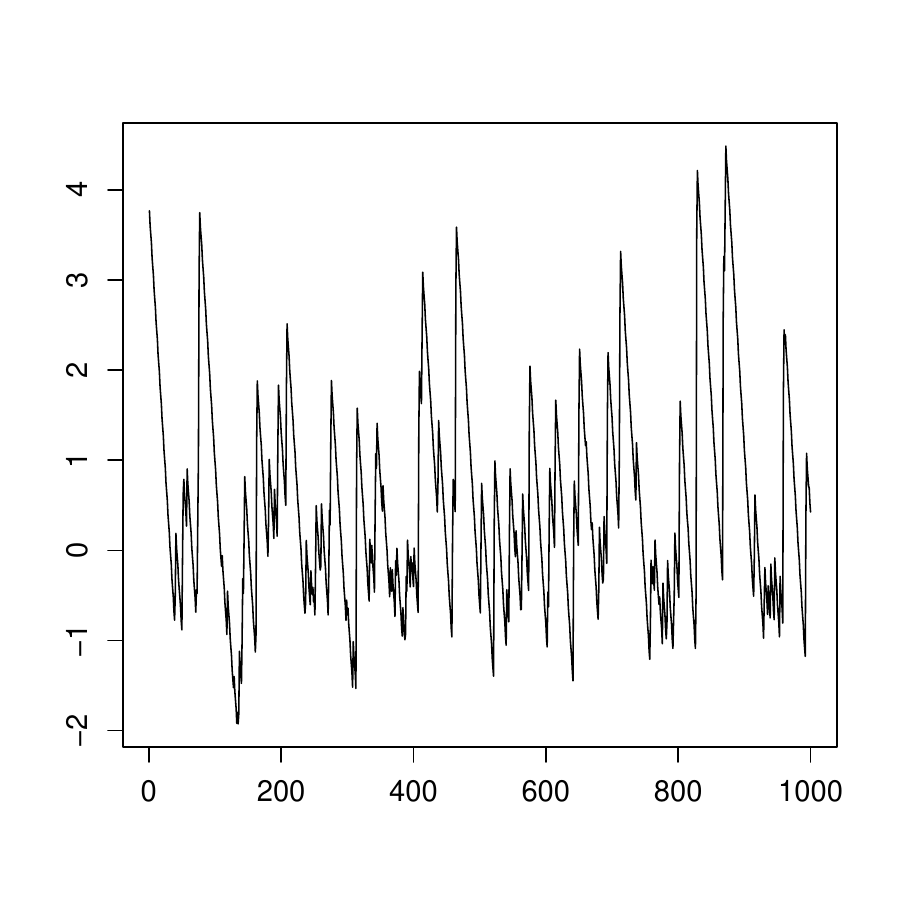}
    \includegraphics[width=0.49\textwidth]{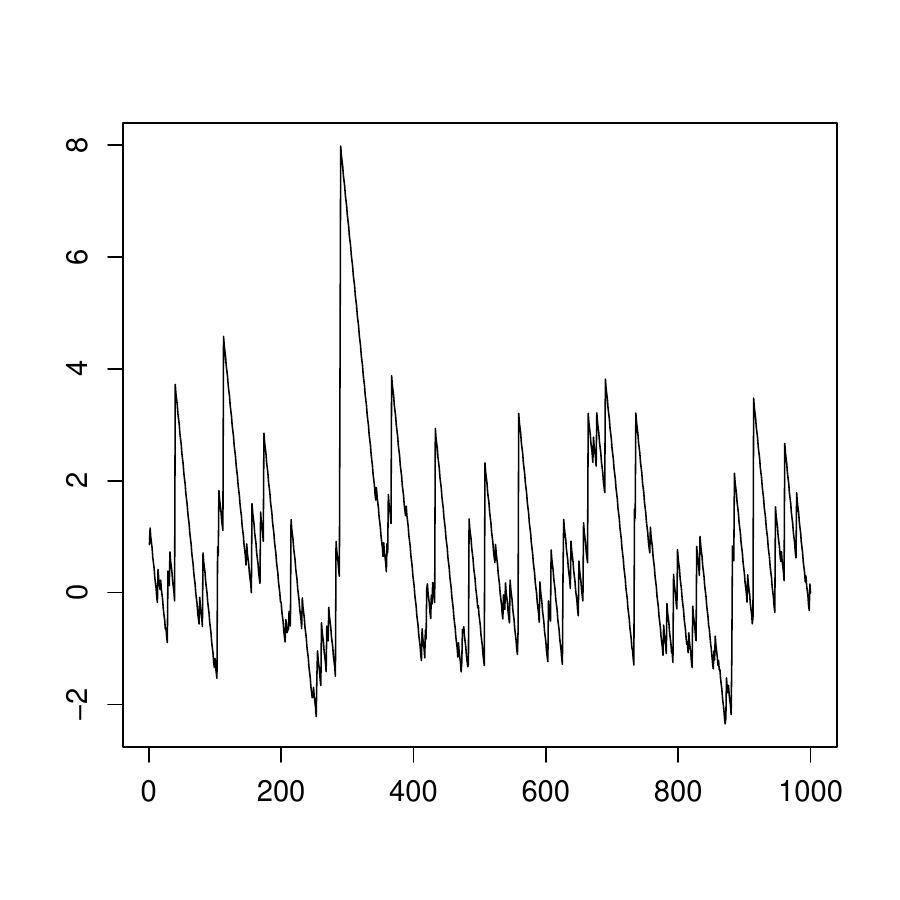}
    \caption{Simulations from stationary Max-ARMA$(p,q)$ processes $\{X_t\}$, presented on Gumbel margins, i.e., for $\log X_t$, with sample sizes $n=1000$:  $(p,q)=(3,0)$ (top row) and $(p,q)=(3,3)$ (bottom row) with parameters $\bm\alpha=(0.85,0.77,0.7)$ (top left), $\bm\alpha=(0.3,0,0.1)$ (top right), $\bm\alpha=(0.85,0.77,0.7)$ and  $\bm\beta=(2,1,0.9)$ (bottom left), and $\bm\alpha=(0.85,0.77,0.7)$ and $\bm\beta=(50,10,5)$ (bottom right).}
    \label{Max-ARMA:fig:simulations}
\end{figure}

To gain further insight into temporal trajectories of series 1-4, we simulated realisations from each series, with Figure~\ref{Max-ARMA:fig:simulations} showing a segment of 1000 consecutive values from each of these processes (after burn-in). To avoid the largest events dominating the image, each series presented in Figure~\ref{Max-ARMA:fig:simulations} is on a standard Gumbel marginal scale, i.e., we show the time series of $\log X_t$. For all four series the dominating feature of the plots is the repeated spikes - sudden jumps up in values - followed by an approximate linear decay (an exponential decay on unit Fr\'echet margins becomes linear on this Gumbel marginal scale). Despite all four series having identical choices of $p$ and, for series 1, 3 and 4, similar values of $\theta$, their trajectories differ: series 2 has more sporadic behaviour with frequent spikes that instantly decay, whereas series~4 has larger jumps from a typical value to an extreme event due to the $\bm\beta$ parameters being much larger than for the other series.

Simulation from these Max-ARMA processes also provides an independent assessment of the theoretically derived properties of $\theta$ and $\chi_{\kappa}$. To obtain Monte Carlo estimates with limited noise we use simulations of length $10^6$ and estimate $\theta(u)$ and $\chi_{\kappa}(u)$ using empirical values based on expressions~\eqref{Max-ARMA:eqn:OBrien} and~\eqref{Max-ARMA:eqn:ChiTau}. These values are given in Table~\ref{Max-ARMA:tab:sim_results}. For estimates of $\theta(u)$, we have $u_n=u$ equal to the 0.95-marginal quantile and $p_n=1$ for all series. Empirical estimates of $\chi_\kappa(u)$ decrease as $\kappa$ increases so the choice of $p_n=1$ is sufficient. Table~\ref{Max-ARMA:tab:sim_results} shows an excellent agreement between these Monte Carlo estimates and the true values for all four series, despite our threshold not being very extreme.

%% file: Sections/inference.tex
\subsection{Model parameterisation}\label{Max-ARMA:subsec:model_param}

In Section~\ref{Max-ARMA:subsec::definition} we show that the parameters $\bm\alpha$ and $\bm\beta$ of a Max-ARMA$(p,q)$ process need to satisfy the conditions of Remarks~\ref{Max-ARMA:rmk::gamma} and~\ref{Max-ARMA:rmk:identifiable} to achieve stationarity and identifiability, respectively, with the conditions of Remark~\ref{Max-ARMA:rmk:identifiable} on the $\bm\alpha$ parameters expressed more parsimoniously in Proposition~\ref{Max-ARMA:prop:identifiability}. In particular, if for any $k=2,\ldots ,p$, we have that $\alpha_1, \ldots ,\alpha_{k-1}$
all satisfy the identifiability conditions of Remark~\ref{Max-ARMA:rmk:identifiable}, then if 
$\alpha_k$ is less than or equal to the lower bound of the constraint~\eqref{Max-ARMA:eqn:alphalowerbound} then  $\alpha_k$ plays no role in determining the sample path of the Max-ARMA process, so it is equivalent to setting $\alpha_k=0$. Similarly, if for any $j=1,\ldots,q$ we have $\beta_j\le \alpha_j$  then $\beta_j$ also does not affect the sample paths, so has no difference from taking $\beta_j=0$. However, for a Max-ARMA$(p,q)$ process to be well-defined we need both the $\alpha_p$ and $\beta_q$ terms to be identifiable, and for statistical inference we cannot allow for multiple points in the parameter space to give processes with identical sample paths if they have identical realisations of the innovation sequence $\{Z_t\}$. Thus, for stationary processes, without any redundancy for lack of identifiability, and for $(p,q)$ to both be uniquely defined, we must further restrict the parameter space, identified in Section~\ref{Max-ARMA:sec:properties}, to the parameter space $\Theta_{p,q}$ given by 
\begin{align*}
    \Theta_{p,q} = \bigg\{ &(\bm\alpha,\bm\beta): 
    0\le \alpha_1<1, 
    \max_{j=1,\ldots ,\lfloor i/2\rfloor} \{\alpha_j\alpha_{i-j}\} \; \leq \; \alpha_i 
    < \; 1    \;\forall\; i=2,\ldots,(p-1), \\
    & \max_{j=1,\ldots ,\lfloor p/2\rfloor} \{\alpha_j\alpha_{p-j}\} \; < \; \alpha_p 
    < \; 1,\beta_0=1, \; \beta_j\geq\alpha_j 
    \;\forall j=1,\ldots,\min\{p,q-1\},\\ & 
    \text{ and for }  q>p, \; \beta_j \; \geq \; 0 \text{ for } j=(p+1),\ldots,(q-1), \; \beta_q \; > \; 0,\\ 
    & \text{ whilst for } p \; \ge \; q, \;
    \beta_q \; > \;\max\{\alpha_q,0\}\bigg\}. 
\end{align*}

This novel formulation for $\Theta_{p,q}$ has the benefit of the parameter space being continuous whilst allowing for any combination of the parameters $\bm\alpha_{-p}$ for $p\geq2$ (since $\alpha_1$ is always identifiable) and $\bm\beta_{-(0,q)}$ (since $\beta_0=1$ is always identifiable) not to affect the sample paths; this is achieved when they satisfy the equal conditions in their respective bounds in the specification of $\Theta_{p,q}$. Imposing the parameter space $\Theta_{p,q}$ on the Max-ARMA$(p,q)$ process has no effect on either $\gamma$, $\theta$ or $\chi_{\tau}$ in expressions~\eqref{Max-ARMA:eq::gamma},~\eqref{Max-ARMA:eq:theta} and~\eqref{Max-ARMA:eq:chi} respectively, and is key for inferences in Sections~\ref{Max-ARMA:subsec:inference} and~\ref{Max-ARMA:sec:Thames}.

Due to the complexity of the parameter space defined above, we reparameterise $\bm\alpha$ and $\bm\beta$ to achieve a more orthogonal parameter space so that inference is easier. Consider the parameters  $\bm\delta=(\delta_1, \ldots ,\delta_p)$ and $\bm\epsilon=(\epsilon_1, \ldots ,\epsilon_q)$ defined by \begin{equation}
    \delta_i = \begin{cases}
        \alpha_1 \quad & \text{for } i=1 \\
        \alpha_i - \max_{j=1,\ldots ,\lfloor i/2\rfloor} \{\alpha_j\alpha_{i-j}\} \quad & \text{for }i=2,\ldots,p,
    \end{cases}\label{Max-ARMA:eqn:delta_def}
\end{equation}
and \begin{equation}
\epsilon_j = \begin{cases} 
        \beta_j - \alpha_j \quad\quad &\text{for } j=1,\ldots,\min\{p,q\} \\ 
        \beta_j \quad\quad &\text{for } j=(p+1),\ldots,q \text{ if }q>p.
\end{cases}\label{Max-ARMA:eqn:epsilon_def}  \end{equation} 
Thus, when $\delta_i=0$, for any $i=1,\ldots ,p-1$ (or when 
$\epsilon_j=0$ for any $j=1,\ldots,q-1$) then $\alpha_i$ (or $\beta_j$) has no impact on the sample path of the Max-ARMA$(p,q)$ process. With this transformation, the new parameter space becomes \begin{equation}
    \tilde\Theta_{p,q} = \big\{ 0\leq \delta_i <\Delta_i \text{ for } i=1,\ldots,(p-1), \; 0< \delta_p <\Delta_i, \; \epsilon_j\geq 0 \text{ for }j=1,\ldots,(q-1), \; \epsilon_q>0 \big\},
    \label{Max-ARMA:eqn:transformed parameterspace}
\end{equation} 
where $\Delta_i$ expresses the bound $\alpha_i<1-\max_{j=1,\ldots ,\lfloor i/2\rfloor} \{\alpha_j\alpha_{i-j}\}$ in terms of $(\delta_1, \ldots ,\delta_{i-1})$. Although $\Delta_i$ is complex for a general $i$, the condition is easily checked after transforming $\bm\delta$ into $\bm\alpha$ and is simple when $p$ is small, e.g., $\Delta_1=1, \Delta_2=1-\delta_1^2$. Though at first sight $\tilde{\Theta}_{p,q}$ may not seem much simpler than $\Theta_{p,q}$, in practice it is much easier to use in optimisations such as in Section~\ref{Max-ARMA:subsec:inference}. This is because the constraints between the $i$th components of the parameters $\bm\alpha$ and $\bm\beta$ have been removed for $i=1, \ldots ,\min\{p,q\}$, as has the complex inequality $\max_{j=1,\ldots ,\lfloor i/2\rfloor} \{\alpha_j\alpha_{i-j}\} \; < \; \alpha_i$ for all $i=2, \ldots ,p$. The new constraint with upper bound $\Delta_i$ is typically satisfied for most trial values of $\bm\delta$ in an optimisation as it would be unlikely for the true parameters to be very close to the upper limits on $\alpha_i$, i.e. corresponding to non-stationarity. Thus we found that using $(\bm\delta,\bm\epsilon)\in \tilde{\Theta}_{p,q}$ is a major simplification to using $\Theta_{p,q}$.




\subsection{Inference strategy}
\label{Max-ARMA:subsec:inference}

\cite{davis1989} considered inference issues in theory for a Max-ARMA$(p,q)$ process. They find super-efficient estimators of the $\bm\alpha$ and $\bm\beta$ parameters when $(p,q)$ are known. To illustrate this, it can easily be shown that for a Max-ARMA(1,0) process $\{X_t\}$, if $X_t>u$ for large $u$ and any $t\in\mathbb{N}$, then $X_{t+1}=\alpha X_t$ for some $\alpha\in(0,1)$ with probability tending to 1 as $u\rightarrow{\infty}$. More generally, for a Max-ARMA$(p,q)$ process, they show that if $\alpha_i$ is identifiable, then $\Pr(X_t=\alpha_i X_{t-i})>0$ and that $X_t/X_{t-i}\ge \alpha_i$ for all $i=1,\ldots ,p$; and they find some similar features involving the $\bm\beta$ parameters. So, for a sample of size $n$, they proposed following estimator, for all $i=1\ldots ,p$, \begin{equation*}
    \hat{\alpha}_i = \min_{t=i+1,\ldots,n}\bigg\{\frac{X_t}{X_{t-i}}\bigg\},
\end{equation*} 
with a positive probability that $\hat{\alpha}_i=\alpha_i$, and this probability tending to 1, with a geometric rate, as $n\rightarrow \infty$.

The pseudo-deterministic behaviour of the Max-ARMA$(p,q)$ process will not be observed exactly in practice for any real-world system. So, our inference strategy differs from that of~\cite{davis1989} as we consider the Max-ARMA$(p,q)$ process to be only an approximation to the actual process generating the data (e.g., river flow data) and we believe this model is useful only when the process is in an extreme state. So, we only assume the Max-ARMA model provides an approximate formulation of the process for extreme observations, i.e., when $\max\{X_{t-1}, \ldots ,X_{t-p}\}>u$, for some high threshold $u$.  Given this perspective, likelihood inference is not suitable for fitting a Max-ARMA($p,q$) process of specified orders $p$ and $q$ to observational data as we do not view that the data actually comes from this precise model. Furthermore, we treat $(p,q)$ as unknown, and therefore they also need estimating.

Instead of likelihood inference, we take a moments-based inference approach for fitting the model to the extremes of observational data. Our strategy is motivated by the approach proposed by~\citet{Rodriguez1988} for rainfall models and continued through a series of extensive work, with a recent example being~\citet{Kaczmarska2015}. In this approach, key properties of the process are derived in closed form and then the model parameters are estimated using a method of generalised moments; this minimises a weighted sum of squared differences between empirical properties of interest and the parametric estimates under the model. In the rainfall context, the generalised moments include features of the body of the process, such as correlations, mean length of dry periods and means of aggregated rainfall over different time windows. Here, as extremes are of most interest, we follow a similar approach but use extremal properties of the process to ensure a good fit, such as $\theta$ and $\chi_k$ for $k\in\mathbb{N}$.

To estimate $\bm\alpha$ and $\bm\beta$ for a Max-ARMA($p,q$) process we use $p+q+2$ moments. Let $\hat{M}_m$ and $M_m$ denote the empirical and true extremal dependence measures, respectively, for $m=1,\ldots,(p+q+2)$. Then the moments are formally defined, for the empirical measures, as 
\begin{equation}
    \hat{M}_m = \begin{cases}
        \hat\theta(u) \quad\quad &\text{for }m=1 \\
        \hat\chi_1(u) \quad\quad &\text{for }m=2 \\
        \hat\chi_{T_m}(u) \quad\quad &\text{for } 2<m<(p+q+2) \\
        \hat\chi_T(u) \quad\quad &\text{for }m=p+q+2, \\
    \end{cases}\label{Max-ARMA:eqn:moments_def}
\end{equation} 
where $\hat\theta(u)$ and $\hat\chi_\kappa(u)$ are  estimates of the sub-asymptotic empirical estimates for a threshold $u$, given in Sections~\ref{Max-ARMA:subsec::extremalindex} and~\ref{Max-ARMA:subsec::chi}, respectively. Furthermore, $T\ge p+q$, is chosen as the value $\kappa$ for which the rate of decay of $\hat{\chi}_{\kappa}(u)$ first changes (i.e., decelerates), and then $T_m=\lfloor T(m-2)/(p+q) \rfloor$ ensures that the lags of $\hat{\chi}_{T_m}(u)$ used are equally spaced for $2\le m\le p+q+2$. In this way, we use values of $\chi_\kappa$ with not necessarily consecutive lags and spread over all lags with the strongest temporal dependence. The true measures $M_m$ are defined analogously, with the parametric forms for extremal dependence measures $\theta$ and $\chi_\kappa$ for a Max-ARMA$(p,q)$ process derived in Propositions~\ref{Max-ARMA:prop::extremalindex} and~\ref{Max-ARMA:prop::chi}, respectively. 

As there is a strong dependence between the estimates of $\chi_{\kappa}$ for different $\kappa$ values, we found it necessary to add additional features to improve model fit. We combine the simple estimators for $\alpha_i \;(i=1,\ldots ,p)$ of~\cite{davis1989} with our inference strategy, exploiting this information into our generalised moment structure as an additional moment to $M_m\;(m=1, \ldots, p+q+2)$,  but using only extreme observation pairs, i.e., when $\min\{X_t,X_{t-i}\}>u$. By combining it with the other generalised moments, we can let our estimate of $\alpha_i$ be the same as the \cite{davis1989} estimator, if that were restricted to a fit on only the largest values, but only if the fit to the other generalised moments is not compromised.

As discussed in Section~\ref{Max-ARMA:subsec:model_param}, we reparametrise $(\bm\alpha,\bm\beta)$ in terms of $(\bm\delta,\bm\epsilon)$ to simplify the parameter space we are working with. However, the extremal dependence measures derived in Propositions~\ref{Max-ARMA:prop::extremalindex} and~\ref{Max-ARMA:prop::chi}, that are required for the method of moments, are defined in terms of $\bm\alpha$ and $\bm\beta$. It is straightforward to obtain expressions for these in terms of the new parameters $\bm\delta$ and $\bm\epsilon$ using the inverse of expressions~\eqref{Max-ARMA:eqn:delta_def} and~\eqref{Max-ARMA:eqn:epsilon_def}, respectively. We minimise the objective function, 
\begin{equation}
    \mathcal{M}(\bm\delta,\bm\epsilon; p,q) = \frac{\omega}{p+q+2}\sum_{m=1}^{p+q+2}(\hat{M}_m-M_m)^2 + \frac{1-\omega}{p}\sum_{i=1}^{p}\min_{t\in T(u,i)}\bigg\{\bigg(\frac{X_{t}}{X_{t-i}}-\alpha_i\bigg)^2 \bigg\},\label{Max-ARMA:eqn:obj_function}
\end{equation}
where $T(u,i)=\{t=1, \ldots,n: \min\{X_t, X_{t-i}\}>u\}$, over $(\bm\delta,\bm\epsilon)\in \tilde{\Theta}_{p,q}$ for fixed $(p,q)$, where $\tilde{\Theta}_{p,q}$ is defined by the set~\eqref{Max-ARMA:eqn:transformed parameterspace}. Here we have weighting terms $\omega/(p+q+2)$ and $(1-\omega)/p$ for $0<\omega<1$, so that as $(p,q)$ are changed $\mathcal{M}(\bm\delta,\bm\epsilon;p,q)$ should be reasonably stable as the two different terms are averaged over their values for $(m,i)$ respectively, and that greater importance is given to the extremal generalised moments as $\omega$ is increased. 

For a given pair $(p,q)$, minimising $\mathcal{M}(\bm\delta,\bm\epsilon;p,q)$ gives our estimated values $(\hat{\bm\delta}_{p,q},\hat{\bm\epsilon}_{p,q})$ of $(\bm\delta,\bm\epsilon)$ and equivalent of $(\bm\alpha,\bm\beta)$. However, we cannot minimise $\mathcal{M}(\hat{\bm\delta}_{p,q},\hat{\bm\epsilon}_{p,q};p,q)$ over $(p,q)$ to find the best values for these indices. By our construction of $\mathcal{M}(\bm\delta,\bm\epsilon;p,q)$ to use averages in the objective function, over $(p+q+2)$- and $p$-terms respectively, we should see a form of stability in the values of $\mathcal{M}(\bm\delta,\bm\epsilon;p,q)$ once $p\ge p_{0}$ and $q\ge q_{0}$ for true values $(p_{0}, q_{0})$, but for smaller values of $(p,q)$, increasing either should result in $\mathcal{M}(\hat{\bm\delta}_{p,q},\hat{\bm\epsilon}_{p,q};p,q)$ decreasing. This suggests a form of elbow plot for order selection, as used in many areas, such as in determining $k$ in a $k$-mean clustering algorithm~\citep{syakur2018}.

\subsection{Marginal inference}\label{Max-ARMA:subsec:marg_inf}
Since we define Max-ARMA($p,q$) processes on unit Fr\'echet margins, we require an additional transformation from the observed stationary process $\{Y_t\}$ to the stationary Max-ARMA($p,q$) process $\{X_t\}$ for inference. We assume that the marginal upper tail of $\{Y_t\}$ has heavy tailed margins, since we believe that any process for which the Max-ARMA process will be suitable to model its dependence structure will have heavy tails, and this is the case for the River Thames data that we analyse in Section~\ref{Max-ARMA:sec:Thames}.

We model the upper tail of the distribution $\{Y_t\}$, above a marginal threshold $u_M$, using a Pareto distribution where $u_M$ is within the sample of observed $\{Y_t\}$ values. 
Specifically, we have that the survivor function is modelled by
\begin{equation}
    \Pr(Y_t>y)= d\left(\frac{u_M}{y}\right)^{c} 
    \mbox{ for }y \ge u_M,
    \label{Max-ARMA:eqn:Pareto}
\end{equation} 
for $c\in\mathbb{R}_+$ and $d\in (0,1)$. To estimate $(c,d)$ we use maximum likelihood methods, making the working assumption that the observations $\{Y_t>u_M\}$ are independent, which gives the estimates 
\begin{equation*}
    \hat{c}=\bigg(\frac{1}{n_u}\sum_{j=1}^{n_u}\log\frac{y_{j}}{u_M}\bigg)^{-1} \quad \text{ and } \quad \hat{d}=\frac{n_u}{n},
\end{equation*} 
where $y_1,\ldots,y_{n_u}$ are the $n_u$ observations of $Y_t$ that exceed $u_M$. 
Here $\hat{c}$ is identical to the Hill estimator~\citep{hill1975}, corresponding to the reciprocal of the mean of the exceedances of $u_M$ on a log scale, and $\hat{d}$ is the sample proportion above $u_M$. Below the threshold $u_M$ we have no theoretical basis for the form of the distribution of $\{Y_t\}$, so we estimate the distribution function
function $F_Y(y)$, for $y< u_M$, using the empirical distribution of $\{Y_t\}$.
As defined, the two components of the estimated distribution for $\{Y_t\}$, denoted as a combined function $\hat{F}_Y$, are continuous at $u_M$.

We use the probability integral transform to map from $\{Y_t\}$ to $\{X_t\}$. Specifically, we use the transformation $X_t=T(Y_t)$ for all $t$, where $T(y)=-1/\log(\hat{F}_Y(y))$. Likewise, after simulating realisations $\{X^*_t\}$ of the fitted Max-ARMA model, we back-transform to give simulated realisations on the original margins, denoted by $\{Y^*_t\}$, where $Y^*_t=\hat{F}^{-1}_Y(\exp(-1/X^*_t))$. Here $\hat{F}_Y^{-1}(v)$ is well-defined for $d\le v<1$ from inverting expression~\eqref{Max-ARMA:eqn:Pareto}, whereas for $0<v<d$ we linearly interpolate $\hat{F}_Y$ between jumps to obtain a one-to-one function with a unique inverse.

%% file: Sections/illustration.tex
We illustrate the inference procedure set out in Section~\ref{Max-ARMA:sec:inf} for fitting a Max-ARMA$(p,q)$ process and selecting ($p,q$) for the UK River Thames daily maximum river flow data, for which four winter segments of the series are shown in Figure~\ref{Max-ARMA:fig:riverflow}. We apply a greater weight to the generalised extremal moments component than that of~\cite{davis1989} parameter estimation by setting $\omega=(p+q+2)/(2p+q+2)$ so that when rescaled by the number of moments in each component, the weights are equal; see expression~\eqref{Max-ARMA:eqn:obj_function}. The extremal moments $\theta$ and $\chi_\kappa$, derived in Propositions~\ref{Max-ARMA:prop::extremalindex} and~\ref{Max-ARMA:prop::chi}, respectively, are defined in terms of an infinite sum which we find is reasonably approximated by a sum up to 100, as discussed in Section~\ref{Max-ARMA:sec:simulation}. 

\begin{figure}
    \centering
    \includegraphics[width=0.45\textwidth]{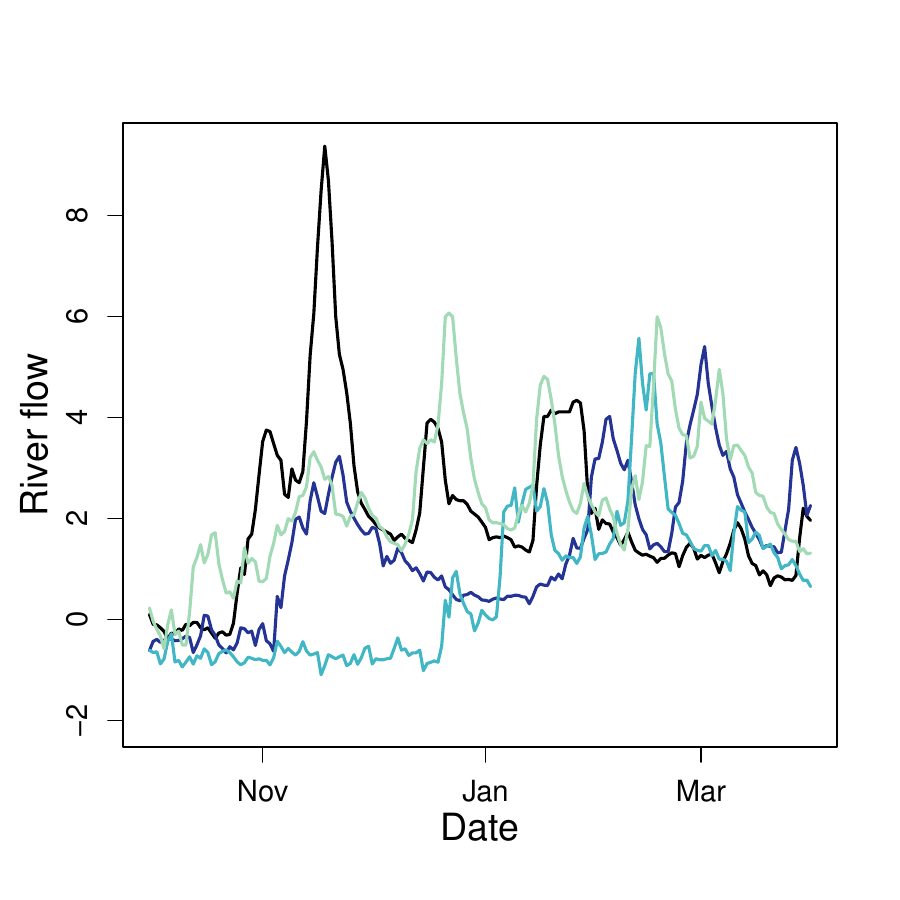}\includegraphics[width=0.45\textwidth]{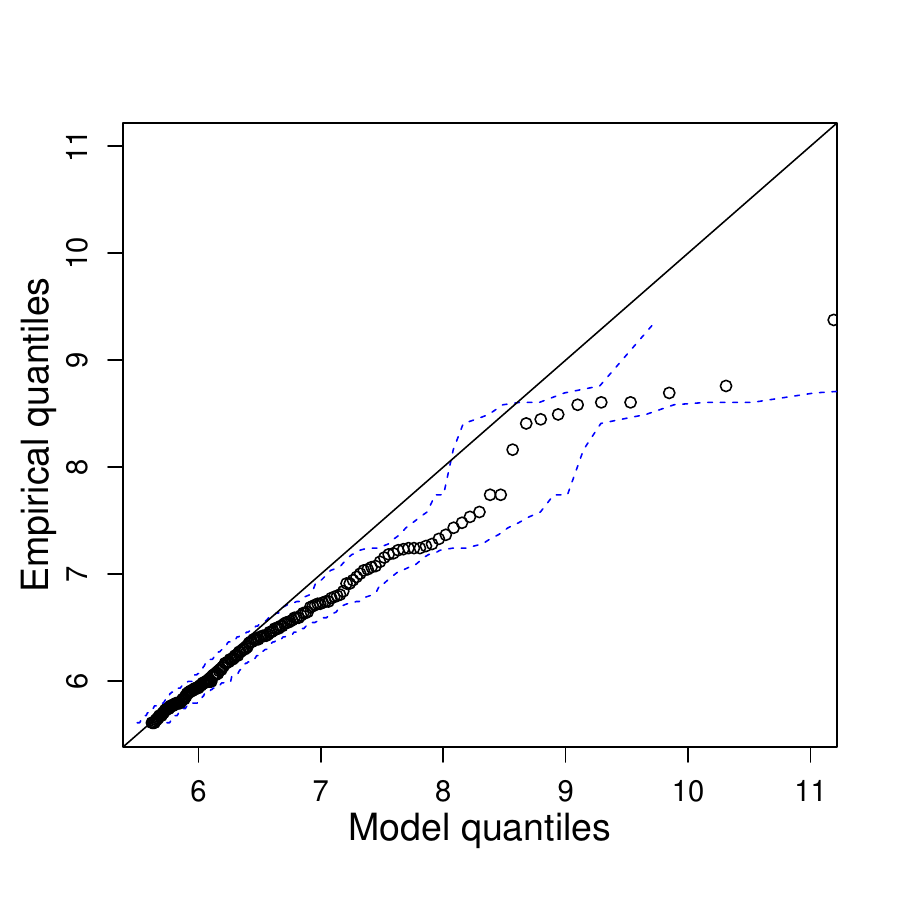}
    \caption{Left: River flow trace plot of the River Thames on Gumbel margins for the winter season (October - March) in 1894/95 (black), 1927/28 (dark blue), 1973/74 (light blue) and 2019/2020 (green). Right: QQ plot of the marginal Pareto tail model fitted to the River Thames exceedances of $u_M$ on Gumbel margins with 95\% tolerance bounds.}
    \label{Max-ARMA:fig:traceplot_frechetmargins}
\end{figure}

Before fitting the Max-ARMA model, which has unit Fr\'echet margins, the observed river flow series $\{Y_t\}$ must be transformed componentwise to also have unit Fr\'echet margins. We do so using the Hill estimator~\citep{hill1975} and probability integral transform, as outlined in Section~\ref{Max-ARMA:subsec:marg_inf}, giving the series $\{X_t\}$. We use a threshold $u_M=270.4$ (equivalent to the $0.98$ quantile) and obtain a Hill estimate of $\hat c=5.1$, so the data has much lighter tails than a unit Fr\'echet series where $c=1$. Figure~\ref{Max-ARMA:fig:traceplot_frechetmargins} shows a QQ plot for the data exceedances of $u_M$, after using the fitted Pareto tail model to transform the data to Gumbel margins, i.e., the same as $\log X_t$. The plot shows evidence that our marginal model fits reasonably well in the upper tail, especially for the less extreme values but with a slight overestimation for the largest observations; changing the threshold $u_M$ could improve the fit. The log transformation of the series is also shown in Figure~\ref{Max-ARMA:fig:traceplot_frechetmargins} (i.e., on Gumbel margins) for the same winter seasons as given in Figure~\ref{Max-ARMA:fig:riverflow}, so that the two largest winter events and two randomly selected winter periods are shown. The data are now on the same marginal scale as our simulations for various Max-ARMA$(p,q)$ processes shown in Figure~\ref{Max-ARMA:fig:simulations}, where the data and simulations have a broadly similar structure in terms of spikes and decays. The largest increase from a typical event to an extreme event is similar to that of simulated series 4, suggesting a large $\beta_j$ parameter, for some $1\leq j \leq q$, might be required to capture a change of this magnitude. The observed data in Figure~\ref{Max-ARMA:fig:traceplot_frechetmargins} has stronger dependence than these simulated series which can be seen by the longer decays following a spike. Additionally, the spikes in our simulations occur instantly, going from a typical value to an extreme in 1 or 2 time lags, whilst the observed data takes much longer to reach a spike due to the large catchment size of the River Thames of $\sim$16,200km$^2$. 

To obtain the moments of expression~\eqref{Max-ARMA:eqn:moments_def} for inference of the river flow series we must find the value of $T$ (the maximum lag $\kappa$ for $\chi_\kappa$ that we use) where the estimate of $\chi_T(u)$ is small and the rate of decay of $\hat\chi_\kappa(u)$ slows down for $\kappa>T$. Figure~\ref{Max-ARMA:fig:riverflow} shows that $T=14$ (i.e., 2 weeks) for both the 0.9 and 0.95 quantile. So, for example, for fitting a Max-ARMA(2,0) we would be interested in extremal moments of $\theta,\chi_1,\chi_{7},\chi_{14}$. 

We fit Max-MARMA($p,q$) models for all combinations of $p=1,2,3$ and $q=0,\ldots,4$ to two sets of the River Thames data: exceedances of the 0.9 and 0.95 quantiles, so that the threshold $u$ in expression~\eqref{Max-ARMA:eqn:obj_function} is set equal to these quantiles. By using these high quantiles, most of the observations we use for inference will be from the winter months when the largest events occur due to the seasonality of river flows. We tested thresholds higher than those shown here but omit the details as the results were noisy so it was difficult to choose $p$ and $q$ in these cases. Thresholds lower than the 0.9 quantile were not considered as we are only interested in extreme events. We compare fits using the minimised objective function values $\mathcal{M}(\hat{\bm\delta}_{p,q},\hat{\bm\epsilon}_{p,q};p,q)$ of expression~\eqref{Max-ARMA:eqn:obj_function} for each $(p,q)$ combination, as well as assessing how close empirical estimates of the extremal dependence measures $\theta$ and $\chi_\kappa$ from the data are to the model based estimates for different orders $(p,q)$.

\begin{figure}
    \centering
    \includegraphics[width=0.45\textwidth]{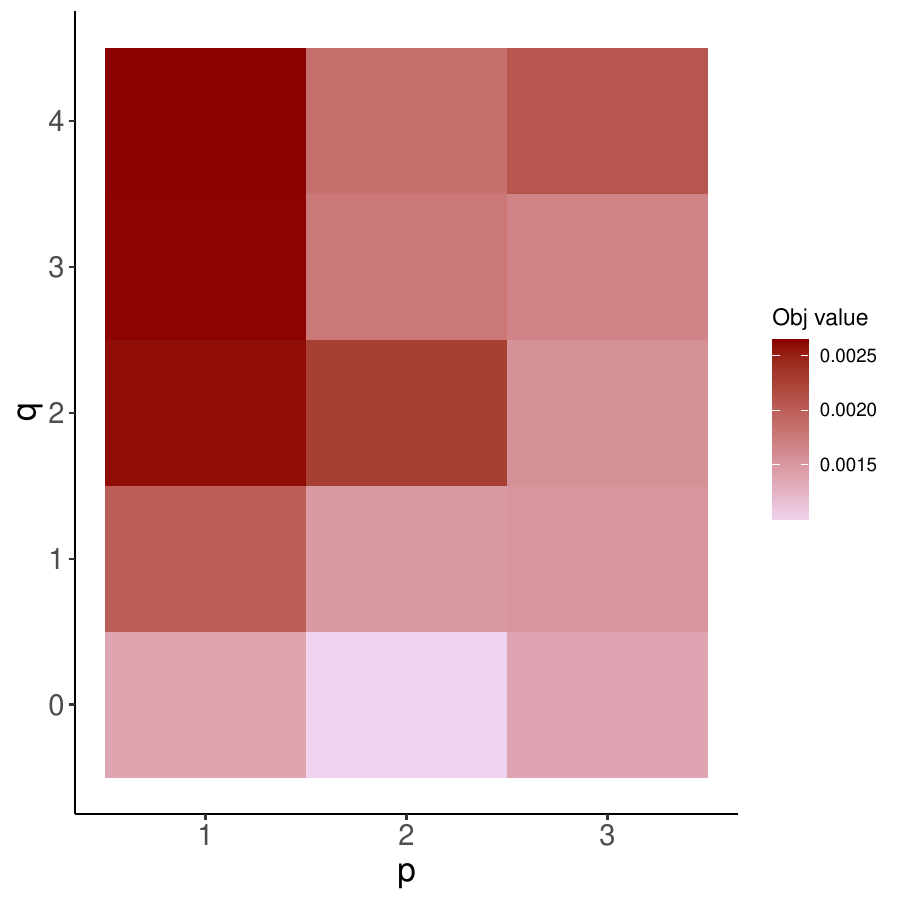}    \includegraphics[width=0.45\textwidth]{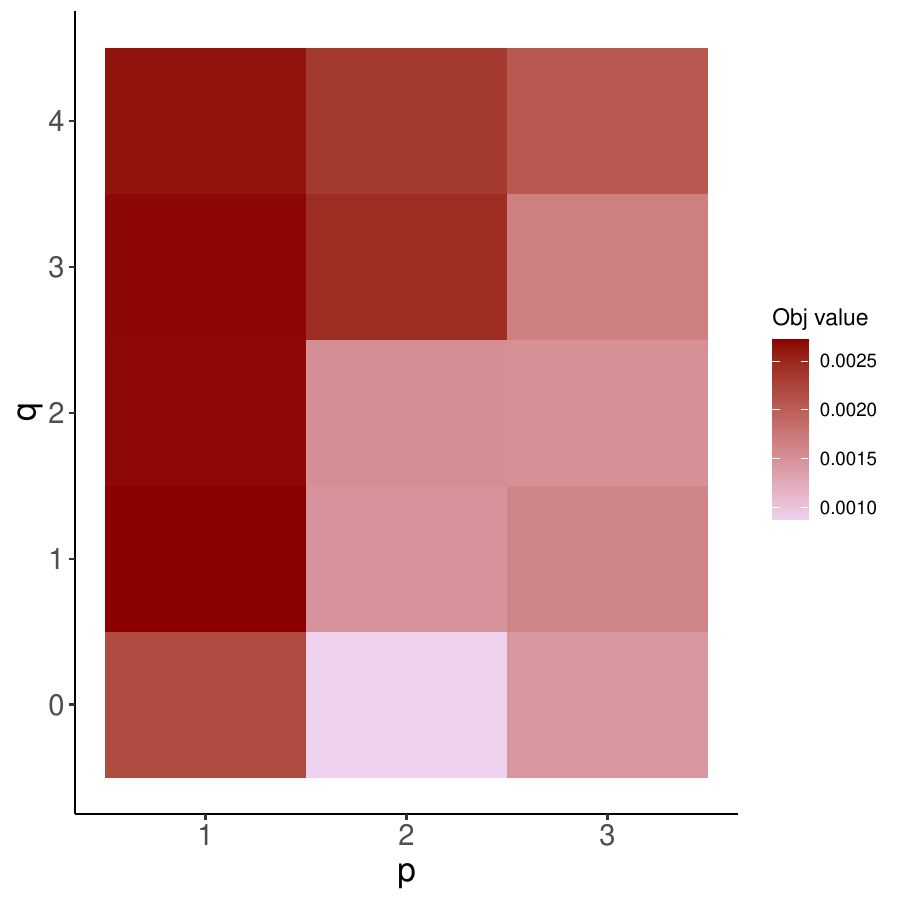}
    \caption{The minimised objective function value $\mathcal{M}(\hat{\bm\delta}_{p,q},\hat{\bm\epsilon}_{p,q};p,q)$ for our moments-based inference of expression~\eqref{Max-ARMA:eqn:obj_function} for Max-ARMA fits of different orders $p=1,2,3$ ($x$-axis) and $q=0,\ldots,4$ ($y$-axis) to the River Thames data using a threshold $u$ of the 0.9 (left) and 0.95 (right) quantiles. Darker (lighter) red indicates a higher (lower) objective function value.}
    \label{Max-ARMA:fig:obj_values}
\end{figure}

Figure~\ref{Max-ARMA:fig:obj_values} shows the minimised $\mathcal{M}(\hat{\bm\delta}_{p,q},\hat{\bm\epsilon}_{p,q};p,q)$ of expression~\eqref{Max-ARMA:eqn:obj_function} for different orders $p$ and $q$, when the model is fitted to exceedances of the 0.9 and 0.95 quantile separately. As mentioned in Section~\ref{Max-ARMA:subsec:inference}, we cannot find the best model fit over ($p,q$) by choosing this to be the $(p,q)$ combination where $\mathcal{M}(\hat{\bm\delta}_{p,q},\hat{\bm\epsilon}_{p,q};p,q)$ is minimised, instead we look for $(p_0, q_0)$ such that there is stability in the $\mathcal{M}(\hat{\bm\delta}_{p,q},\hat{\bm\epsilon}_{p,q};p,q)$ values for all $p\geq p_0$ and $q\geq q_0$. For the model fits to exceedances of the 0.9 quantile, we observe stability in the minimised $\mathcal{M}(\hat{\bm\delta}_{p,q},\hat{\bm\epsilon}_{p,q};p,q)$ for $(p,q)\geq(2,3)$. For both choices of threshold $u$, $\mathcal{M}(\hat{\bm\delta}_{p,q},\hat{\bm\epsilon}_{p,q};p,q)$ is relatively stable across all values of $q$ for a fixed $p$ when $p=3$. For exceedances of the 0.95 quantile, $\mathcal{M}(\hat{\bm\delta}_{p,q},\hat{\bm\epsilon}_{p,q};p,q)$ appears to be stable for $q\geq3$ across all $p=1,2,3$, suggesting a Max-ARMA($3,3$) is the best fitting model based on this criteria, but we make further comparisons below for choosing our selected values of $(p,q)$ here.

In Figures~\ref{Max-ARMA:fig:sims_extremaldep} and~\ref{Max-ARMA:fig:final_3_3} we study estimates of the extremal dependence measures from a given Max-ARMA model across models with different orders ($p,q$). We use simulation methods to evaluate our model based estimates rather than using the theoretical limits derived in Section~\ref{Max-ARMA:sec:ext_measures} that we used in the objective function~\eqref{Max-ARMA:eqn:obj_function}. This is because we are interested in the sub-asymptotic extremal dependence measures $\theta(u)$ and $\chi_\kappa(u)$ so that we can compare these with the corresponding estimates from the observed series. To find these model based estimates, we simulate a large sample and obtain empirical estimates from this sample from the fitted Max-ARMA model using Monte Carlo methods, where we limit the Monte Carlo noise by taking the simulated sample to be of length $10^6$. 

First, we compare empirical estimates of extremal dependence measures $\theta(u)$ and $\chi_\kappa(u)$ for $\kappa=1,7,14$ from the observed data and model-based estimates, when fit to exceedances of the 0.95 quantile. These are shown in Figure~\ref{Max-ARMA:fig:sims_extremaldep}. The Max-ARMA models with orders $(2,0)$, $(2,1)$, $(3,0)$, $(3,2)$ and $(3,3)$ have estimates of $\theta(u)$ that lie within the 95\% confidence intervals (found via bootstrapping) of the corresponding empirical estimate for the data. These same models, as well as when $p=q=2$, also give model estimates of $\chi_1(u)$ within the confidence intervals (based on the Binomial sampling distribution) of the associated empirical estimates. Models with $p=1$ always underestimate $\theta(u)$ compared to the empirical estimates. All models overestimate $\chi_7(u)$, except for $p=3$ and $q=4$, whilst most models (except when $(p,q)$ is $(1,0)$, $(2,4)$ or $(3,4)$) give $\chi_{14}(u)$ estimates that lie within the 95\% confidence intervals from the empirical estimates. Therefore, combining the results from Figures~\ref{Max-ARMA:fig:obj_values} and~\ref{Max-ARMA:fig:sims_extremaldep}, we conclude that a Max-ARMA($3,3$) is the best fitting model for river flow exceedances of the 0.95 quantile at the River Thames. For this model we obtain parameter estimates $\hat{\bm\alpha}=(0.69, 0.78, 0.54)$ and $\hat{\bm\beta}=(3.15, 2.16, 0.99)$. Since all $\beta_j>\alpha_j$ for $j=1,2,3$, all $\bm\beta$ are identifiable and affect the sample path of simulations from this model. Additionally, $\alpha_2>\alpha_1^2$ whilst {$\alpha_3-\alpha_1\alpha_2>0$, but small, so primarily $\alpha_1$ and $\alpha_2$ influence the sample paths of the simulations from this model.

\begin{figure}
    \centering
    \includegraphics[width=0.45\textwidth]{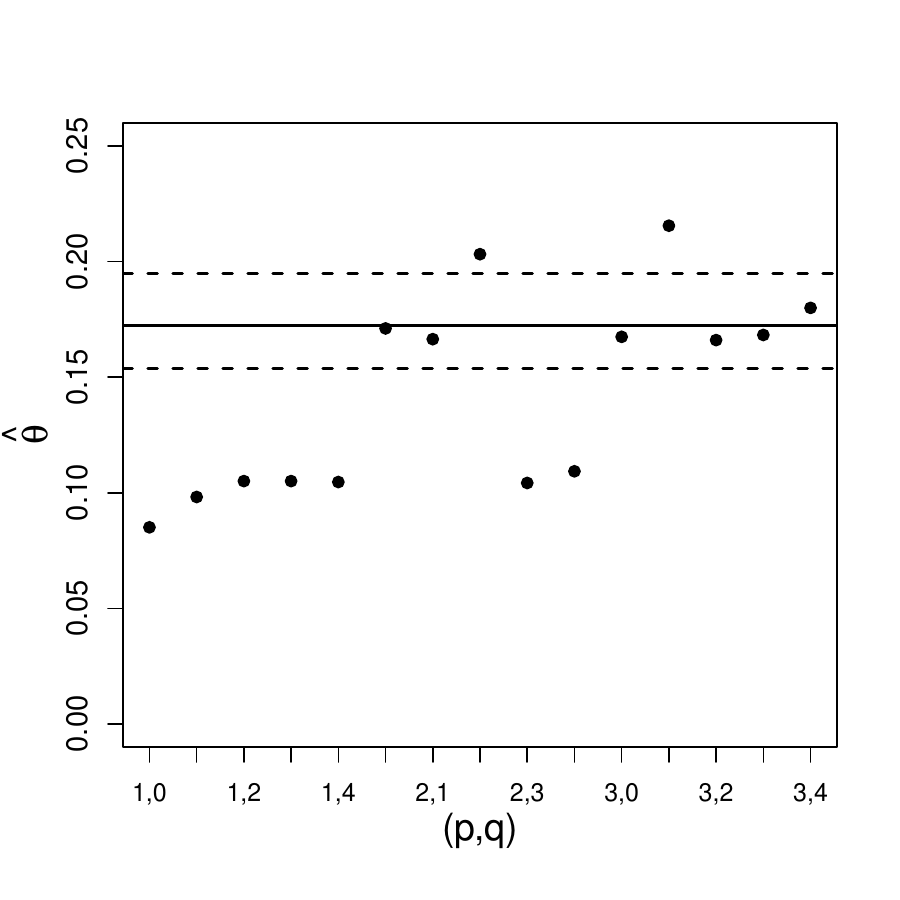}    \includegraphics[width=0.45\textwidth]{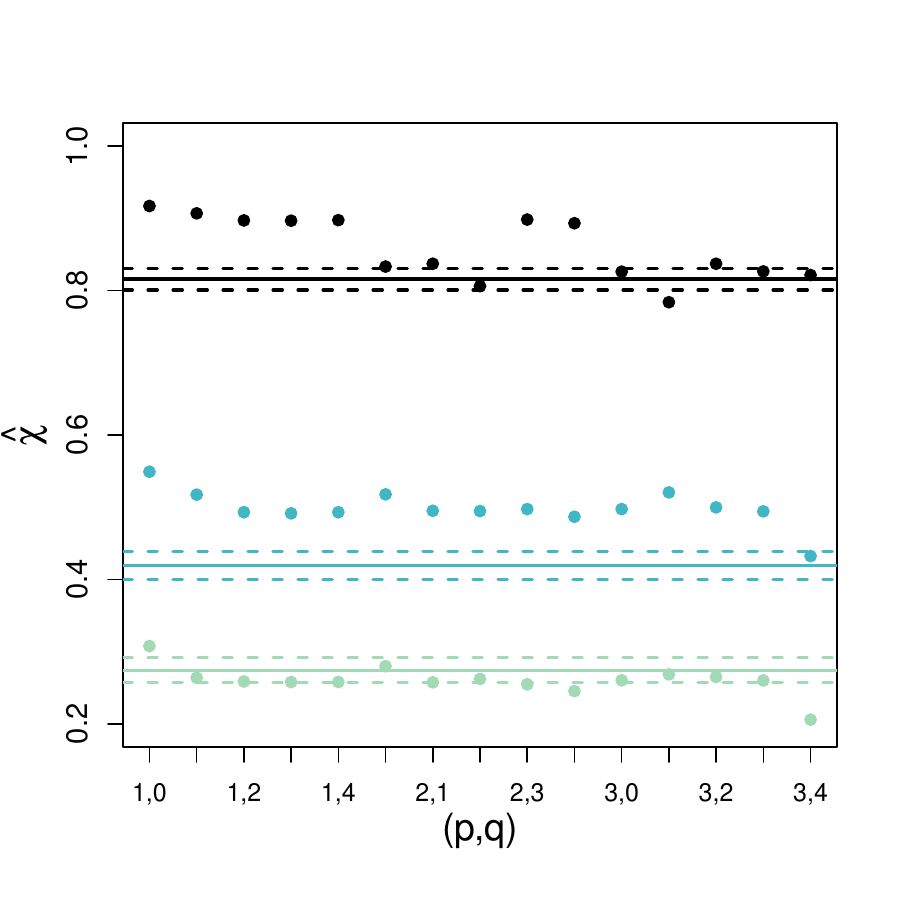}
    \caption{Estimates of $\theta(u)$ (left) and $\chi_\kappa(u)$ (right) for $k=1,7$ and $14$ (black, blue and green, respectively) using a threshold $u$ of the 0.95 quantile for the River Thames with empirical estimates (solid lines) and estimates using fitted Max-ARMA models (points) with varying orders $(p,q)$ ($x$-axis). Horizontal dashed lines show 95\% confidence intervals for the empirical estimates of $\theta(u)$ and $\chi_\kappa(u)$ for the data.}
    \label{Max-ARMA:fig:sims_extremaldep}
\end{figure}

Figure~\ref{Max-ARMA:fig:final_3_3} shows estimates of $\chi_\kappa(u)$ for $\kappa=1,\ldots,14$ from the data, with 95\% confidence intervals, and from the Max-ARMA(3,3) model based estimates. For $\kappa=1$ and $\kappa\geq11$, we capture the structure of data estimate of $\hat{\chi}_\kappa(u)$ very well, but for $2\leq\kappa\leq10$ the model based estimates slightly overestimate this asymptotic dependence measure. For the other choices of $(p,q)$ we tried, we did not obtain model based estimates of $\chi_\kappa(u)$ within the confidence intervals of the data estimates for lags in the range $2\leq\kappa\leq 10$.

Figure~\ref{Max-ARMA:fig:final_3_3} shows four simulated time series plots over 183 days (corresponding to the length of the winter seasons shown in Figure~\ref{Max-ARMA:fig:riverflow})  using our fitted Max-ARMA(3,3) model, with the simulated series transformed onto the original data scale using the probability integral transform, as discussed in Section~\ref{Max-ARMA:subsec:marg_inf}. We show time series from the two largest simulated events (where the simulated time series is the same length as the total River Thames winter data, so these events should be comparable in size to the largest observed events) as well as two randomly selected time series segments to illustrate \textit{typical} behaviour. All four simulated time series segments exhibit a saw-tooth structure because $\hat\alpha_2>\hat\alpha_1$ and we find this behaviour in all model fits with $p>1$. Our simulated realisations of time series exhibit similar behaviour to the observed data in the decay, following a major spike, so our model is capturing the features of decay of the original series well. The simulations of the typical level segments of the time series also exhibit a similar number of smaller peaks to the observed data in Figure~\ref{Max-ARMA:fig:riverflow}. A noticeable feature of the observed data that our simulations fail to capture is that the major spikes can take several days (or even up to a month, as for the flood event of winter 1894; see black time series segment in Figure~\ref{Max-ARMA:fig:riverflow}) to rise to their peak, however, in the simulated realisations of the fitted models the spikes occur almost instantly. So this is a limitation of the best fitting model for the data that we have considered. 

\begin{figure}
    \centering
    \includegraphics[width=0.45\textwidth]{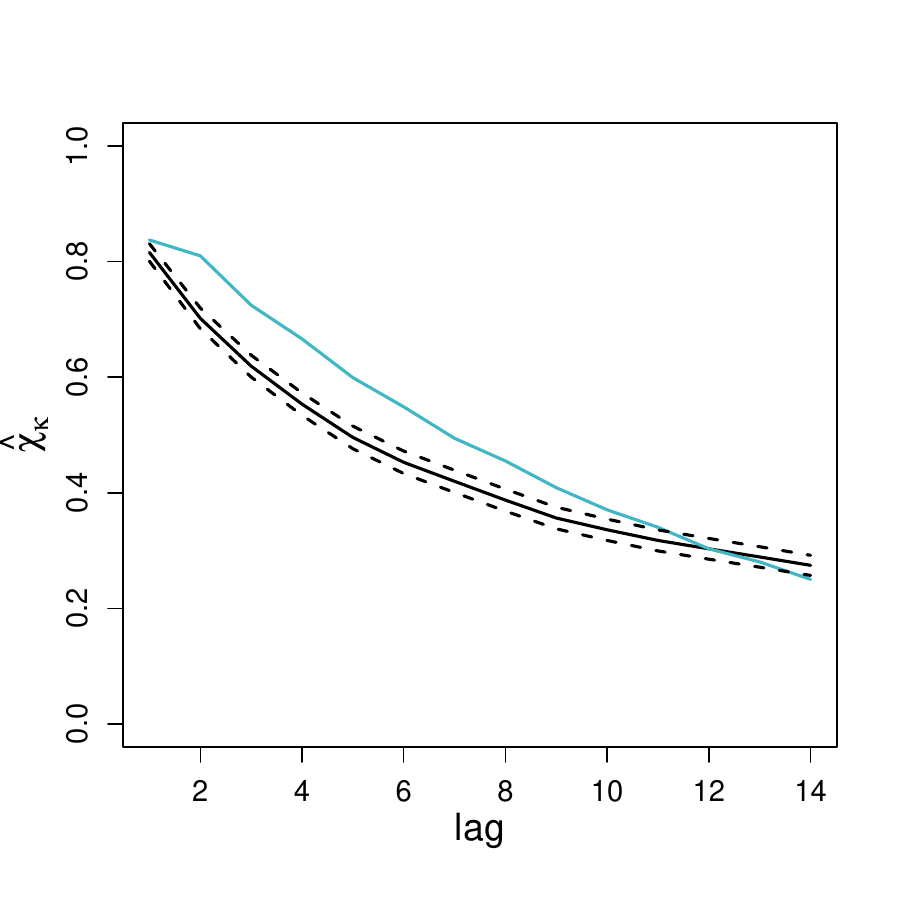}
    \includegraphics[width=0.45\textwidth]{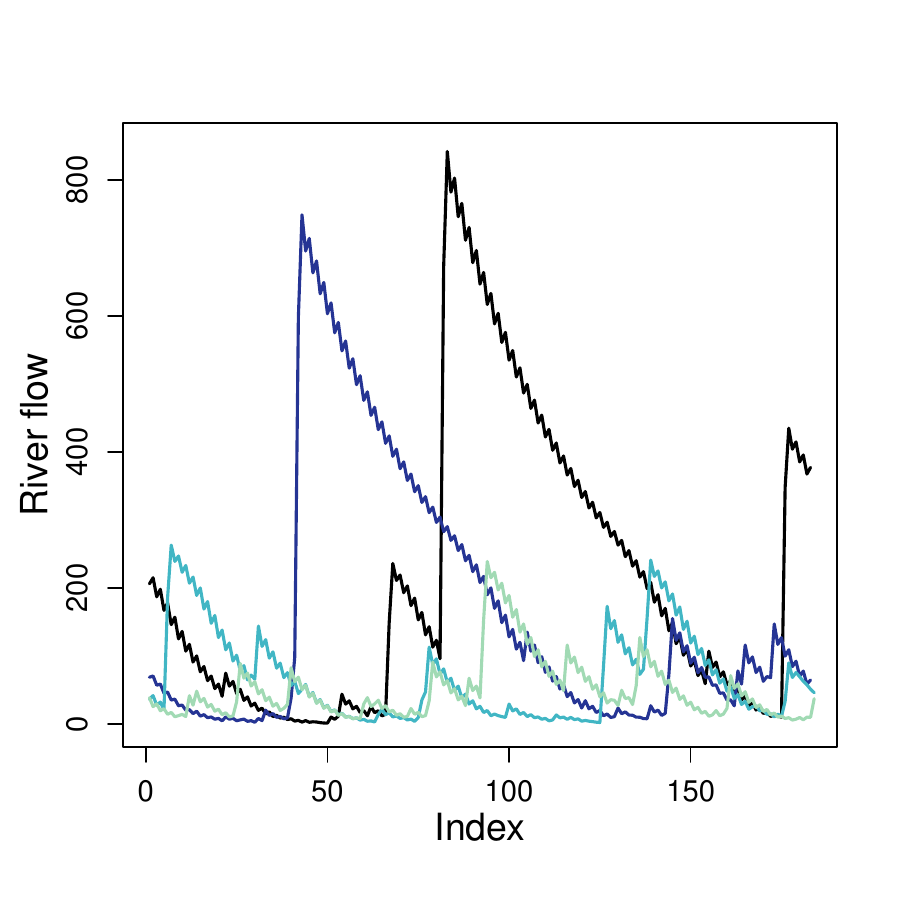}
    \caption{Left: Empirical estimates of $\chi_\kappa(u)$ for $\kappa=1,\ldots,14$, with $u$ equal to the 0.95 quantile, for the data (black with 95\% confidence intervals in dashed lines) and estimates from a Max-ARMA(3,3) model fit to the River Thames data (blue). Right: Four time series plots of length 183 (corresponding to the length of the winter season October-March) from the fitted Max-ARMA(3,3) to exceedances of the 0.95 quantile of the River Thames data, transformed to the original data scale. The time series plots correspond to the two largest events (black and dark blue) and two randomly selected series (light blue and green).}
    \label{Max-ARMA:fig:final_3_3}
\end{figure}

Our results are primarily illustrative for the inference procedure, rather than to demonstrate a definitive approach for choosing $(p,q)$. We consider this to be of potential interest for future work, but for our purposes the choices are sufficient for barrier closure assessment, see the discussion below. However, to get a better fit of the model for the spike segments of the time series we could have imposed a much larger value for $q$, with the $\beta_j$ increasing with $j$ so that for a large innovation given by $Z_t$ say, this would lead to a rising spike in the series through successive values with $\{X_{t+j}= \beta_j Z_t\}$ for $j=0, \ldots ,q$, which would grow towards a spike at time $q$. 
The instant spike behaviour we have estimated with $q=3$ is typical of river flows with a smaller area and non-chalk based catchments areas, where these types of rivers are known in the hydrology area as {\it flashy},  
see \cite{stewart2008flood}. So it would appear that, with a more careful choice of $q$, we have the flexibility with the Max-ARMA model to describe well the extremal features and key hydrology aspects of the time series profiles of rivers in different catchment sizes. Further work could investigate an approach for selecting the optimal $(p,q)$ orders for these data or explore applying our inference procedure to rivers with smaller catchments.

From the perspective of the problem of barrier closure assessment for the River Thames, as the set out in Section~\ref{Max-ARMA:intro}, we have demonstrated the capability of a Max-ARMA model coupled with a Pareto marginal tail model to capture the core aspects of extremal dependence structure in river flows of the River Thames, i.e., the marginal tail decay, $\theta$ and $\chi_\kappa$ which primarily describe the magnitude and duration of the events more than the actual profile of events. It is these dependence aspects that are crucial for forecasting future closure rates of the Thames Barrier; as barrier closures are determined by 
peak sea levels per the tidal cycle (i.e., peak tides plus skew surge) and/or river flows in extreme states. Specifically, as extreme skew surge events tend to last 1-3 days yet extreme river levels can last up to 14-20 days with a mean cluster length of 5 days, it is possible that more than one extreme skew surge event can occur during the duration of an extreme river flow event. Accounting for this possibility is vital for determining barrier closure rates and their clustering in time. So long term simulations from these models provide good approximations to the observed series for considering barrier closure rate properties, as well as allowing for more extreme events than have been observed both in terms of their marginal sizes and their temporal duration.
Given that river flow is independent of skew surge~\citep{SvenssonJones2004}, future work can look at combining simulations of both variables (using the skew surge model of~\citep{DArcy2023}, the deterministic predicted peak tide series, and the Max-ARMA model for river flow) to estimate future closure rates, and how they cluster in time, and ultimately understand the reliability of the barrier as we face unprecedented challenges resulting from anthropogenic climate change. 

%% file: Sections/proofs.tex
\subsection{Proof of Proposition~\ref{Max-ARMA:prop:identifiability}}
\begin{proof}
The proof works by induction. Let $P_p$ be the statement:  All $\alpha_i>0$ for $i=1,\ldots,k$ where $k\leq p$ are identifiable if $\max_{i=1,\ldots ,\lfloor k/2\rfloor} \{\alpha_i\alpha_{k-i}\} \leq \alpha_k$ for all $k\leq p$.

For $k\geq2$, assume that all statements $\{P_i\}_{i=2}^k$ hold. Then consider statement $P_{k+1}$. For $\alpha_{k+1}$ to be identifiable, we know from Remark~\ref{Max-ARMA:rmk:identifiable} that \begin{equation}
    \max_{\mathcal{R}_{k+1}}\bigg\{\prod_{\substack{i=1,\ldots,k:\\\alpha_i>0}}\alpha_i^{a_i} \bigg\} < \alpha_{k+1}. \label{Max-ARMA:eq:rmk2_in_proof}
\end{equation}
This can be written as the maximum of the following components, \begin{equation}
    \max\left\{\alpha_1\max_{\mathcal{R}_{k}}\bigg\{\prod_{\substack{i=1,\ldots,(k-1):\\\alpha_i>0}}\alpha_i^{a_i} \bigg\},
    \max_{\mathcal{R}_{k+1}}\bigg\{\prod_{\substack{i=2,\ldots,k:\\\alpha_i>0}}\alpha_i^{a_i} \bigg\}\right\}< \alpha_{k+1}, \label{Max-ARMA:eq:iden_a1_factored}
\end{equation} where the first term above is any elements in the maxima set of expression~\eqref{Max-ARMA:eq:rmk2_in_proof} that contain $\alpha_1$, whilst the second term is those without $\alpha_1$ terms. 

Since $\alpha_k$ is identifiable, we know from Remark~\ref{Max-ARMA:rmk:identifiable} that, \begin{equation*}
        \max_{\mathcal{R}_{k}}\bigg\{\prod_{\substack{i=1,\ldots,(k-1):\\\alpha_i>0}}\alpha_i^{a_i} \bigg\} < \alpha_{k}. 
\end{equation*} Then inequality~\eqref{Max-ARMA:eq:iden_a1_factored} becomes \begin{equation*}
    \max\left\{ \alpha_1\alpha_k, \max_{\mathcal{R}_{k+1}}\bigg\{\prod_{\substack{i=2,\ldots,k:\\\alpha_i>0}}\alpha_i^{a_i} \bigg\} \right\} < \alpha_{k+1}.
\end{equation*}

\noindent Similarly, we rewrite the second term so that the above inequality becomes \begin{equation*}
   \max\left\{ \alpha_1\alpha_k, 
  \alpha_2\max_{\mathcal{R}_{k-1}}\bigg\{\prod_{\substack{i=1,\ldots,(k-2):\\\alpha_i>0}}\alpha_i^{a_i} \bigg\},
  \max_{\mathcal{R}_{k+1}}\bigg\{\prod_{\substack{i=3,\ldots,k:\\\alpha_i>0}}\alpha_i^{a_i} \bigg\}\right\} < \alpha_{k+1},
\end{equation*} where the second term is any elements containing $\alpha_2$ and the third term is those without $\alpha_2$. Since $\alpha_{k-1}$ is identifiable, we can rewrite this using Remark~\ref{Max-ARMA:rmk:identifiable} as 
\begin{equation*}
    \max\left\{\max\{\alpha_1\alpha_k, \alpha_2\alpha_{k-1}\},
    \max_{\mathcal{R}_{k+1}}\bigg\{\prod_{\substack{i=3,\ldots,k:\\\alpha_i>0}}\alpha_i^{a_i} \bigg\}\right\} < \alpha_{k+1}. 
\end{equation*} Continuing in this way for a further $(\delta-2)$ iterations where $\delta\in\mathbb{N}$ and $\delta>2$, we obtain \begin{equation}
   \max\left\{ \max\{\alpha_1\alpha_k, \alpha_2\alpha_{k-1},\ldots,\alpha_{\delta}\alpha_{k-\delta}\}, 
   \max_{\mathcal{R}_{k+1}}\bigg\{\prod_{\substack{i=(\delta+1),\ldots,k:\\\alpha_i>0}}\alpha_i^{a_i} \bigg\}\right\} < \alpha_{k+1}. \label{Max-ARMA:eq:maxterm_general}
\end{equation}


\noindent Consider when $\delta=\lfloor k/2 \rfloor$, then \begin{equation*}
    \max_{\mathcal{R}_{k+1}}\bigg\{\prod_{\substack{i=\lfloor k/2 \rfloor+1,\ldots,k:\\\alpha_i>0}}\alpha_i^{a_i} \bigg\} = \emptyset,
\end{equation*} because $i=1,\ldots,\lfloor k/2 \rfloor -1$ are required to meet the conditions of the set $\mathcal{R}_{k+1}$, but we only consider $i=(\delta+1),\ldots,k$ and for these values, there are no solutions in $a_i\in(0,\ldots,k+1)$ such that $\sum_{i=\lfloor k/2 \rfloor}^{p} ia_i = k+1$. For $\delta > \lfloor k/2 \rfloor$, this is always the case and so the above set is always empty. This is the first time when the second term in expression~\eqref{Max-ARMA:eq:maxterm_general} becomes the empty set, because when $\delta=\lfloor k/2 \rfloor - 1$, this term becomes \begin{equation*}
    \max_{\mathcal{R}_{k+1}}\bigg\{\prod_{\substack{i=\lfloor k/2 \rfloor,\ldots,k:\\\alpha_i>0}}\alpha_i^{a_i} \bigg\} = \alpha_{\lfloor k/2 \rfloor}\alpha_{(k+1) - \lfloor k/2 \rfloor}.
\end{equation*}
Therefore, $\delta=\lfloor k/2 \rfloor$, giving the required result,
    $\max\{\alpha_1\alpha_k, \alpha_2\alpha_{k-1}, \ldots, \alpha_{\lfloor k/2 \rfloor}\alpha_{k-\lfloor k/2 \rfloor}\} \leq \alpha_{k+1}$.
\end{proof}

\subsection{Proof of Proposition~\ref{Max-ARMA:prop::extremalindex}}
\begin{proof}

Let $\{X_t\}$ be a stationary Max-ARMA($p,q$) process, defined in Section~\ref{Max-ARMA:subsec::definition}, under the conditions of~\cite{davis1989} (see Remark~\ref{Max-ARMA:rmk::gamma}). Consider the maximum of the process $\{X_t\}$, first under an IID variables assumption (denote this sequence by $\{\hat{X}_t\}$) with an identical marginal distribution to that of $\{X_{t}\}$. Let $\hat{M_n}=\max\{\hat{X_1},\ldots,\hat{X_n}\}$ be the maximum and the limiting non-degenerate distribution of the normalised $\hat{M}_n$ is denoted by $\hat{G}$. We find the distribution of the scaled maximum by exploiting this independence assumption, \begin{equation*}
    \Pr(\hat M_n/n\leq x)=\exp(-n/nx)=\exp(-1/x):=\hat G(x) \quad \text{for } x>0.
\end{equation*}
Next, we derive the same limiting distribution for the Max-ARMA process, i.e., without the independence assumption; denote this by $G(x)$. We begin by considering the case when $n=3$ before considering the $n>q$ case, as the former reveals the key steps in the latter. 
 

\paragraph{Case $n=3$:}
By the definition of a Max-ARMA($p,q$)  we can write  \begin{align*}
    M_3                 &= \max\{X_1, \alpha_1 X_1, \beta_0 Z_2, \beta_1 Z_1, \alpha_1 X_2, \alpha_2 X_1, Z_3, \beta_1 \beta_0 Z_{2}, \beta_2 Z_1 \} \\
                        &= \max\{X_1, \alpha_1X_1, \beta_0 Z_2, \beta_1Z_1, \alpha_1^2 X_1, \alpha_1\beta_0 Z_2, \alpha_1\beta_1 Z_1,\alpha_2X_1, \beta_0Z_3, \beta_1Z_2, \beta_2Z_1\} \\
                        &= \max\big\{\max\{1,\alpha_1,\alpha_2\}X_1, \max\{\beta_1, \alpha_1\beta_1, \beta_2\}Z_1, \max\{\alpha_1\beta_0, \beta_0, \beta_1\}Z_2, \beta_0 Z_3\big\}.
\end{align*} As $0<\alpha_{i}<1$ for all $i=1,\ldots,p$ under stationarity and $\beta_0=1$, from Remark~\ref{Max-ARMA:rmk::gamma}, we have\begin{equation*}
    \Pr(M_3 \leq nx) = \Pr(X_1\leq nx, \max\{\beta_1,\beta_2\}Z_1 \leq nx, \max\{\beta_0,\beta_1\}Z_2\leq nx, Z_3 \leq nx).
\end{equation*} Since $Z_2$ and $Z_3$ are independent of $X_1$ and $Z_1$, this joint probability can be factorised to give
\begin{equation*}
    \Pr(M_3\leq nx) = \Pr(X_1\leq nx, \max\{\beta_1,\beta_2\}Z_1 \leq nx)\Pr(\max\{\beta_
    0,\beta_1\}Z_2\leq nx)\Pr(Z_3 \leq nx).
\end{equation*} We simplify the joint probability of $X_1$ and $Z_1$ by
\begin{align*}
    \Pr(X_1 \leq nx, \max\{\beta_1,\beta_2\}Z_1 \leq nx) &= \Pr(X_1\leq nx)\Pr(\max\{\beta_1,\beta_2\} Z_1 \leq nx|X_1\leq nx)\\&=\Pr(X_1\leq nx)\Pr(\max\{\beta_1,\beta_2\} Z_1 \leq nx|Z_1\leq nx),
\end{align*} where the conditioning event $\{X_1\leq nx\}$ changes to $\{Z_1\leq nx\}$ as this is the only information from the event involving $X_1$ that is relevant to $\{\max\{\beta_1,\beta_2\} Z_1\leq nx\}$. It follows that,
\begin{align*}
    \Pr(M_3\leq nx) = &\Pr(X_1\leq nx)\Pr(Z_3\leq nx)\Pr(Z_2\leq nx/\max\{
    \beta_0,\beta_1\})\\ &\times \min\{\exp(-\gamma(\max\{\beta_1,\beta_2\}-1)/(nx)), 1\},
\end{align*} because if $\max\{\beta_1,\beta_2\} \leq 1$, then $\Pr(\max\{\beta_1,\beta_2\} Z_1 \leq nx|Z_1\leq nx)=1$ but otherwise, this conditional probability would need to be evaluated. This final term can be written more simply in our last expression for $M_3$ as \begin{align*}
    \Pr(M_3\leq nx) = &\Pr(X_1\leq nx)\Pr(Z_3\leq nx)\Pr(Z_2\leq nx/\max\{
    \beta_0,\beta_1\})\\ &\times \exp\Big(-\frac{\gamma}{nx}(\max\{\beta_0,\beta_1,\beta_2\}-1)\Big).
\end{align*} 


\paragraph{Case $n>q$:} 
Following the same logic as for the $n=3$ case, using the notation $\beta_M=\max\{\beta_0, \ldots ,\beta_q\}$ which, by definition of the Max-ARMA process, satisfies $\beta_M\ge 1$ and $\beta_{M:j} = \max\{\beta_0,\beta_1,\ldots,\beta_j\}$ so that $1\leq\beta_{M:j}\leq\beta_M$ for all $j=0,1,\ldots,q$ and $\beta_{M:q}=\beta_M$, the distribution of the rescaled maxima is 
\begin{align*}
    \Pr(M_n/n\leq x) = &\Pr(X_1\leq nx)\prod_{j=0}^{q-1} \Pr\big(Z_{n-j} \leq nx/\beta_{M:j}\big)\prod\limits_{i=2}^{n-q}\Pr\big(Z_{i}\leq nx/\beta_M\big)   
    \\ &\times\exp\bigg(-\frac{\gamma}{nx}[\beta_M-1]\bigg).  
\end{align*}  As $X_t\sim\text{Fr\'echet}(1)$ and $Z_t\sim\text{Fr\'echet}(\gamma)$ with $0<\gamma<\infty$ defined in Remark~\ref{Max-ARMA:rmk::gamma}, we  have 
\begin{align*}
     \Pr(M_n/n\leq x) = &\exp\bigg(-\frac{1}{nx}\bigg)\bigg\{\prod_{j=0}^{q-1}\exp\bigg(-\frac{\gamma \beta_{M:j}}{nx}\bigg)\bigg\}\\ &\times\exp\bigg(-\frac{\gamma\beta_M}{x}\bigg[1-\frac{q-2}{n}\bigg]\bigg) \exp\bigg(-\frac{\gamma}{nx}(\beta_M-1)\bigg)\\
    = &\exp\bigg(-\frac{1}{nx}\Big[1+\gamma\sum_{j=0}^{q-1}\beta_{M:j}\Big]\bigg)\exp\bigg(-\frac{\gamma\beta_M}{x}\left[1-\frac{q-2}{n}\right]\bigg)\\&\times\exp\bigg(-\frac{\gamma}{nx}(\beta_M-1)\bigg) \\
    \rightarrow &\;\big[\exp(-1/x)\big]^{\gamma \beta_M} = \big[\hat{G}(x)\big]^{\gamma \beta_M} \quad \text{as }n\rightarrow\infty.
\end{align*}
\noindent This limit follows as the first and last terms tend to 1 and in the second term $(q-2)/n\rightarrow0$. From Section~\ref{Max-ARMA:subsec::extremalindex} the extremal index $\theta$ is defined via $G(x)=[\hat G(x)]^\theta$, hence $\theta=\gamma \beta_M=\gamma\max\{\beta_0,\beta_1,\ldots,\beta_q\}$.

\end{proof}

\subsection{Proof of Proposition~\ref{Max-ARMA:prop::chi}}
\begin{proof}
The definition of $\chi_\kappa$ can be rewritten for a stationary Max-ARMA process $\{X_t\}$ with unit Fr\'echet margins (using conditional probability and a Taylor expansion) as 
$\chi_\kappa=\lim_{x\rightarrow \infty}\chi_\kappa(x)$ with, as $x\rightarrow \infty$,
\begin{align*}
 \chi_\kappa(x) &=\frac{\Pr(X_{t+\kappa}>x,X_{t}>x)}{\Pr(X_{t}>x)}= \frac{\Pr(X_{t+\kappa}>x,X_{t}>x)}{1-\exp(-1/x)} \\
    &= \frac{\Pr(X_{t+\kappa}>x,X_{t}>x)}{1-(1-1/x-\mathcal{O}(x^{-2}))}= x\Pr(X_{t+\kappa}>x,X_{t}>x)[1+o(1)],
\end{align*} 
where the calculations for the denominator 
follow as $X_t\sim $Fre\'chet(1).
So we need to find the probability of the event $\mathcal{J}^X(x,\kappa):=\{X_{t+\kappa}>x,X_{t}>x\}$ for different lags $\kappa$. In this proof, we consider $\kappa\in\mathbb{N}$ only, but $\chi_\kappa=\chi_{-\kappa}$ holds, due to the symmetry of the above expression, so our results hold for  $\kappa\in\mathbb{Z}$.  

To determine the asymptotic behaviour of the joint probability in $\chi_{\kappa}(x)$ we make the following partition, with $\beta_M=\max\{\beta_0, \ldots ,\beta_q\}\geq 1$ and $M^Z_{s:t}:=\max\{Z_{s},\ldots,Z_{t}\}$ for $s\le t$, 
\begin{align*}
    \Pr(\mathcal{J}^X(x,\kappa))
= &\Pr(\mathcal{J}^X(x,\kappa),
M^Z_{t+1:t+\kappa}<x/\beta_M) +  
\Pr(\mathcal{J}^X(x,\kappa),M^Z_{t+1:t+\kappa}\geq x/\beta_M)\\
    = &\Pr(\mathcal{J}^X(x,\kappa),M^Z_{t+1:t+\kappa}
    <x/\beta_M)[1+o(1)],
\end{align*} as $x\rightarrow\infty$. The reason that the probability of the second term in the partition is smaller order than the probability of the first partition term is because it requires the occurrence of an extreme $Z_{j}$ value from at least one of only $\kappa$ values for $j$, which has a probability that is an order of magnitude smaller than not requiring this event.

First, we identify some conditions that control which type of extreme event the process exhibits. For $t\in \mathbb{Z}$ and $\delta\in \mathbb{N}\cup\{0\}$, we define the events
\begin{eqnarray*}
\mathcal{H}^Z(x,\delta,\kappa)
 & = & \{M^Z_{t-q-\delta:t-1-\delta} <x/\beta_M,
\beta_M Z_{t-\delta}>x,
M^Z_{t-\delta+1:t+\kappa} <x/\beta_M\},\\ 
\mathcal{H}^X(x,\delta)
& = & \{M^X_{t-p-\delta:t-\delta-1}<x/\alpha_M\},
\end{eqnarray*}
where $M^X_{s:t}:=\max\{X_{s},\ldots,X_{t}\}$ for $s\le t$ and $\alpha_M=\max\{\alpha_1,\ldots,\alpha_p\}$ which, by definition of the Max-ARMA process, satisfies $\alpha_M>0$. The event $\mathcal{H}^Z(x,\delta,\kappa)$ gives that there is only one extreme value of the innovation process in the $(q+\delta+\kappa)$ values prior to and including at time $(t+\kappa)$. Specifically, at time $(t-\delta)$ the innovation exceeds $x$ but all the other innovations are less than $x$ as $x/\beta_M\le x$. The event $\mathcal{H}^X(x,\delta)$ requires that prior to $(t-\delta)$, all $p$ previous values of the Max-ARMA process are sufficiently small that they cannot produce an extreme value. So if event $\mathcal{H}(x,\delta,\kappa):=\mathcal{H}^Z(x,\delta,\kappa)\cap\mathcal{H}^X(x,\delta)$ occurs then the only feature of the Max-ARMA and innovation process up to $(t-\delta)$ that can produce a value of the Max-ARMA process that exceeds $x$ at time $(t-\delta)$ and subsequently is $Z_{t-\delta}$.


Then we can rewrite the joint probability of interest by partitioning over the time when the large innovation $Z_{t-\delta}$ occurs before time $t$, i.e., for $\delta=\{0,\mathbb{N}\}$. Then we can express the probability as 
\begin{equation}
    \Pr(\mathcal{J}^X(x,\kappa),   
    M^Z_{t+1:t+\kappa}<x/\beta_M) = \sum\limits_{\delta=0}^{\infty}
    \Pr(\mathcal{J}^X(x,\kappa), 
\mathcal{H}(x,\delta,\kappa))
[1+o(1)],
\label{Max-ARMA:eqn::chi_with_delta}
\end{equation} 
as $x\rightarrow \infty$, where the additional inclusion of additional terms of $\mathcal{H}(x,\delta,\kappa)$ on the right hand side only change the probability by a little $o(1)$ term in $x$. We now focus on finding the asymptotic behaviour of a generic term in the sum in expression~\eqref{Max-ARMA:eqn::chi_with_delta}. Specifically, 
\begin{eqnarray}
\Pr(\mathcal{J}^X(x,\kappa), 
\mathcal{H}(x,\delta,\kappa))
 & = &
\Pr(\mathcal{H}(x,\delta,\kappa))
\Pr(\mathcal{J}^X(x,\kappa) \mid
\mathcal{H}(x,\delta,\kappa)).
\label{Max-ARMA:eqn::chi_with_delta_term}
\end{eqnarray} 
The marginal probability here can be written asymptotically as $\beta_M\gamma/x$ as $x\rightarrow \infty$ as the probability of the event $\{\beta_M Z_{t-\delta}>x\}$ has this limiting behaviour and all the rest of the finite events in 
$\mathcal{H}(x,\delta,\kappa)$ have probabilities tending to one.
If event  $\mathcal{H}(x,\delta,\kappa)$ occurred for large enough $x$,
and all $\alpha_i$ and $\beta_j$ coefficients were non-zero, then 
$X_{t-\delta}=Z_{t-\delta}$, $X_{t-\delta+1}=\max\{\alpha_1,\beta_1\}Z_{t-\delta}=\gamma_1 Z_{t-\delta}$, $X_{t-\delta+\tau}=\gamma_{\tau}
Z_{t-\delta}$ for all $\tau\ge 1$, where $\gamma_{\tau}$ is defined by
expression~\eqref{Max-ARMA:eqn:Edefn}. Thus, under this conditioning,  the process is deterministic given the value of $\beta_M Z_{t-\delta}>x$. When any of the Max-ARMA coefficients are zero there is a potential for $\gamma_{\tau}=0$ for some $\tau$, and in these cases the associated $X_{t-\delta+\tau}=o_p(x)$ as $x\rightarrow \infty$. So we then have \begin{eqnarray}
\Pr(\mathcal{J}^X(x,\kappa) \mid
\mathcal{H}(x,\delta,\kappa))
& = & \Pr(\gamma_{\delta}Z_{t-\delta}>x, \gamma_{\delta+\kappa}Z_{t-\delta}>x|\beta_M Z_{t-\delta}>x)[1+o(1)] 
\nonumber \\
& = & 
\Pr(\min\{\gamma_{\delta},\gamma_{\delta+\kappa}\}Z_{t-\delta}>x|\beta_M Z_{t-\delta}>x)[1+o(1)] 
\nonumber \\
& = & 
\Pr(Z_{t-\delta}>x/\min\{\gamma_{\delta},\gamma_{\delta+\kappa}\})/\Pr(\beta_M Z_{t-\delta}>x)[1+o(1)] \nonumber \\
& \rightarrow & 
\min\{\gamma_{\delta},\gamma_{\delta+\kappa}\},
\label{Max-ARMA:eqn::chi_with_delta_term2}
\end{eqnarray} 
as $x\rightarrow \infty$. As expression~\eqref{Max-ARMA:eqn:Edefn} shows, $\max_{\delta\in \mathbb{N}}(\gamma_{\delta})\le \beta_M$, so in the third equality  the event of interest is a subset of the conditioning event. Also, the final limit exploits the property that $\Pr(Z_{t-\delta}>y)\sim \beta_M \gamma/y$ as $y\rightarrow \infty$. 

Combining these results together we have that 
\[
\chi_{\kappa} =\lim_{x\rightarrow \infty} \chi_{\kappa}(x) = 
\lim_{x\rightarrow \infty} \Pr(\mathcal{J}^X(x,\kappa))/\Pr(X_t>x)= 
\gamma\sum\limits_{\delta=0}^{\infty}\min\{\gamma_{\delta},\gamma_{\delta+\kappa}\}.
\]
This sum will always converge because $\gamma_{\tau}\rightarrow0$ as $\tau\rightarrow\infty$ so for large $\tau$ $\min\{\gamma_{\delta}, \gamma_{\delta+\kappa}\}<1$ and each $\gamma_\tau$ geometrically decays for large $\tau$.
\end{proof}

%% file: Sections/discuss.tex
A key feature of all the stationary Max-ARMA$(p,q)$ processes is that they are asymptotically dependent at all lags, 
i.e., $\chi_{\tau}>0$, with this term defined by limt~\eqref{Max-ARMA:eqn:ChiTau}. Thus this class of processes excludes the possibility of asymptotic independence at different lags, i.e., $\chi_{\tau}=0$ for at least some $\tau\ge 1$, see \cite{LedfordTawn2003}, so extensions of 
Max-ARMA$(p,q)$ processes that allow for asymptotic independence are of interest. One such example is 
the power max-autoregressive (pMax-ARMA) process
proposed by \cite{Ferreira2011}.  
This process takes the form 
$X_t=\max\{X_{t-1}^{\alpha_1},Z_t\}$ with $0<\alpha_1<1$ and $Z_t$ an IID sequence of random variables with real positive support. They demonstrate that this process is asymptotically independent for all lags. However, they do not extend this for a general pMax-ARMA($p,q$) process so the identifiability and stationarity constraints, that we address for the Max-ARMA process, are not relevant there; deriving conditions under which these properties hold would be an interesting avenue for further work. It would also be very interesting to develop a class of models that joins well between these two formulations, so that either asymptotic dependence or asymptotic independence can occur at different lags. Perhaps a way to achieve this is to 
change this innovation series to have a more rapidly decaying tail. An alternative is to change the process in
expression~\eqref{Max-ARMA:eq::Max-ARMA_definition} from combining terms using the $L_{\infty}$-norm to instead being the $L_r$-norm, for some $1<r<\infty$, as that provides a natural link between the Max-ARMA process and the $L_1$-norm case, corresponding to the standard ARMA$(p,q)$ processes, with issues like this have been considered by 
\cite{Schlather2001} in a different context.

%% file: main.bbl
\begin{thebibliography}{}

\bibitem[Box and Jenkins, 1970]{BoxJenkins1970}
Box, G.~E. and Jenkins, G.~M. (1970).
\newblock {\em Time Series Analysis: Forecasting and Control}.
\newblock Holden-Day, San Francisco.

\bibitem[Coles et~al., 1999]{Coles1999}
Coles, S.~G., Heffernan, J., and Tawn, J.~A. (1999).
\newblock Dependence measures for extreme value analyses.
\newblock {\em Extremes}, 2(4):339--365.

\bibitem[Dale et~al., 2014]{dale2014}
Dale, M., Wicks, J., Mylne, K., Pappenberger, F., Laeger, S., and Taylor, S. (2014).
\newblock Probabilistic flood forecasting and decision-making: an innovative risk-based approach.
\newblock {\em Natural hazards}, 70:159--172.

\bibitem[Davis and Mikosch, 2009]{DavisMikosch2009}
Davis, R.~A. and Mikosch, T. (2009).
\newblock The extremogram: a correlogram for extreme events.
\newblock {\em Bernoulli}, 15(4):977--1009.

\bibitem[Davis and Resnick, 1989]{davis1989}
Davis, R.~A. and Resnick, S.~I. (1989).
\newblock {Basic properties and prediction of max-ARMA processes}.
\newblock {\em Advances in Applied Probability}, 21(4):781--803.

\bibitem[D’Arcy et~al., 2023]{DArcy2023}
D’Arcy, E., Tawn, J.~A., Joly, A., and Sifnioti, D.~E. (2023).
\newblock Accounting for seasonality in extreme sea-level estimation.
\newblock {\em The Annals of Applied Statistics}, 17(4):3500--3525.

\bibitem[{Environment Agency}, 2021]{ThamesBarrier}
{Environment Agency} (2021).
\newblock {Thames Estuary 2100: 10-year review - technical monitoring report}.
\newblock \url{https://www.gov.uk/government/publications/thames-estuary-2100-te2100-monitoring-reviews/thames-estuary-2100-10-year-monitoring-review-2021}.
\newblock Accessed 14/01/24.

\bibitem[{Environment Agency}, 2023]{EA_river_data}
{Environment Agency} (2023).
\newblock {Daily mean flow (m3/s) time series for Kingston}.
\newblock \url{https://environment.data.gov.uk/hydrology/station/8496ce69-482c-406a-a2f0-ac418ef8f099}.

\bibitem[Ferreira, 2011]{Ferreira2011}
Ferreira, M. (2011).
\newblock On tail dependence: a characterization for first-order max-autoregressive processes.
\newblock {\em Mathematical Notes}, 90:882--893.

\bibitem[Heffernan et~al., 2007]{heffernan2007}
Heffernan, J.~E., Tawn, J.~A., and Zhang, Z. (2007).
\newblock Asymptotically (in) dependent multivariate maxima of moving maxima processes.
\newblock {\em Extremes}, 10:57--82.

\bibitem[Hill, 1975]{hill1975}
Hill, B.~M. (1975).
\newblock A simple general approach to inference about the tail of a distribution.
\newblock {\em The Annals of Statistics}.

\bibitem[Hsing et~al., 1988]{hsing1988exceedance}
Hsing, T., H{\"u}sler, J., and Leadbetter, M.~R. (1988).
\newblock On the exceedance point process for a stationary sequence.
\newblock {\em Probability Theory and Related Fields}, 78(1):97--112.

\bibitem[Kaczmarska et~al., 2015]{Kaczmarska2015}
Kaczmarska, J.~M., Isham, V.~S., and Northrop, P. (2015).
\newblock Local generalised method of moments: an application to point process-based rainfall models.
\newblock {\em Environmetrics}, 26(4):312--325.

\bibitem[Leadbetter et~al., 1983]{Leadbetter1983}
Leadbetter, M., Lindgren, G., and Rootz\'{e}n, H. (1983).
\newblock {\em {Extremes and Related Properties of Random Sequences and Processes}}.
\newblock Springer-Verlag, New York.

\bibitem[Ledford and Tawn, 2003]{LedfordTawn2003}
Ledford, A.~W. and Tawn, J.~A. (2003).
\newblock Diagnostics for dependence within time series extremes.
\newblock {\em Journal of the Royal Statistical Society: Series B}, 65(2):521--543.

\bibitem[Nelsen, 2006]{nelsen2006}
Nelsen, R.~B. (2006).
\newblock {\em {An Introduction to Copulas}}.
\newblock Springer, New York.

\bibitem[O'Brien, 1987]{obrien1987}
O'Brien, G.~L. (1987).
\newblock Extreme values for stationary and {M}arkov sequences.
\newblock {\em The Annals of Probability}, 15(1):281--291.

\bibitem[Robinson and Tawn, 2000]{robinsontawn2000}
Robinson, M.~E. and Tawn, J.~A. (2000).
\newblock Extremal analysis of processes sampled at different frequencies.
\newblock {\em Journal of the Royal Statistical Society, Series B}, 62(1):117--135.

\bibitem[Rodriguez-Iturbe et~al., 1988]{Rodriguez1988}
Rodriguez-Iturbe, I., Cox, D.~R., and Isham, V. (1988).
\newblock A point process model for rainfall: further developments.
\newblock {\em Proceedings of the Royal Society of London. A}, 417(1853):283--298.

\bibitem[Schlather, 2001]{Schlather2001}
Schlather, M. (2001).
\newblock Limit distributions of norms of vectors of positive iid random variables.
\newblock {\em The Annals of Probability}, 29(2):862--881.

\bibitem[Smith and Weissman, 1994]{SmithWeissman1994}
Smith, R.~L. and Weissman, I. (1994).
\newblock Estimating the extremal index.
\newblock {\em Journal of the Royal Statistical Society: Series B}, 56(3):515--528.

\bibitem[Stewart et~al., 2008]{stewart2008flood}
Stewart, E.~J., Kjeldsen, T.~R., Jones, D.~A., and Morris, D.~G. (2008).
\newblock The flood estimation handbook and {UK} practice: past, present and future.
\newblock In {\em Flood Risk Management: Research and Practice}. CRC Press.

\bibitem[Svensson and Jones, 2004]{SvenssonJones2004}
Svensson, C. and Jones, D.~A. (2004).
\newblock Dependence between sea surge, river flow and precipitation in south and west {B}ritain.
\newblock {\em Hydrology and Earth System Sciences}, 8(5):973--992.

\bibitem[Syakur et~al., 2018]{syakur2018}
Syakur, M., Khotimah, B.~K., Rochman, E., and Satoto, B.~D. (2018).
\newblock Integration k-means clustering method and elbow method for identification of the best customer profile cluster.
\newblock {\em IOP Conference Series: Materials Science and Engineering}, 336:012017.

\bibitem[Trace-Kleeberg et~al., 2023]{tracekleeberg2023}
Trace-Kleeberg, S., Haigh, I.~D., Walraven, M., and Gourvenec, S. (2023).
\newblock How should storm surge barrier maintenance strategies be changed in light of sea-level rise? a case study.
\newblock {\em Coastal Engineering}, 184:104336.

\end{thebibliography}
